\documentclass[pdflatex,sn-nature]{sn-jnl}

\usepackage{graphicx}%
\usepackage{multirow}%
\usepackage{amsmath,amssymb,amsfonts}%
\usepackage{amsthm}%
\usepackage{mathrsfs}%
\usepackage[title]{appendix}%
\usepackage{xcolor}%
\usepackage{textcomp}%
\usepackage{manyfoot}%
\usepackage{booktabs}%
\usepackage{algorithm}%
\usepackage{algorithmicx}%
\usepackage{algpseudocode}%
\usepackage{listings}%

\theoremstyle{thmstyleone}%
\theoremstyle{thmstyletwo}%

\theoremstyle{thmstylethree}%
\newcommand{\AuAu} {Au~+~Au }
\begin{document}

\title[Article Title]{Nuclear equation-of-state at high density and multi-messenger astronomy: contribution of heavy-ion collisions}


\author*[1]{\fnm{Arnaud} \sur{Le~F\`{e}vre}}\email{A. LeFevre@gsi.de}



\affil*[1]{\orgdiv{Research Division}, \orgname{GSI Helmholtzzentrum f\"{u}r Schwerionenforschung GmbH}, \orgaddress{\street{Planckstraße 1}, \city{D-64291 Darmstadt}, 
\country{Germany}}}




\abstract{In the past decades, heavy-ion collisions (HIC) at intermediate energies have allowed to probe the nuclear equation-of-state (EoS) of both symmetric and asymmetric nuclear matter over a broad range of densities. In particular, flow has proven to be a powerful observable. Combining the symmetry energy and the symmetric nuclear matter constraints of the EoS from HIC allowed to predict a density dependence of the pressure in a neutron star, up to about 2.5 times saturation density ($n_{sat}$), which agrees with recent astronomical measurements deduced from gravitational waves and pulsar observations. So far, the accuracy from HIC expectations is comparable to the latter up to 1.5 $n_{sat}$. In these studies, a fundamental aspect is the determination of the profile of densities that are probed by experimental observables used to constrain the EoS. In the near future, new experiments like ASY-EOS performed at higher incident energy and with better accuracy will push further the frontier of the knowledge of the symmetry energy at higher density. These efforts cannot be conclusive without a reliable uncertainty determination, which is related to the reliability of transport model dependencies. Improvements and breakthroughs in transport model simulations and nuclear theory are therefore expected in a joint effort towards HIC contributions to the field of neutron-star physics, including the contribution of strangeness and of the QCD phase transition. 
}

\keywords{Nuclear matter equation-of-state, Symmetry energy, Heavy-ion collisions, Collective flows, Particle production, Neutron Stars, Transport models, QCD phase transition, Strangeness}



\maketitle

\section{Introduction}\label{sec:intro}

The investigation of the nuclear matter equation-of-state (EoS) at high densities is an important subject in modern nuclear physics. 
It determines how nuclear matter behaves under conditions that differ from those found inside atomic nuclei and plays a significant role in phenomena 
such as heavy-ion collisions, neutron-star mergers, the determination of neutron-star mass–radius relations, and the modelling of supernova explosions. 
Our limited knowledge of the EoS is largely due to the challenges involved in solving the many-body problem when realistic nuclear interactions are considered.
Over the past decades, both theoretical and experimental studies have led to substantial progress in describing the EoS for isospin-symmetric matter 
as well as for isospin-asymmetric matter (commonly referred to as the symmetry energy), 
particularly at densities above the saturation density (usually denoted $n_{sat}$ or $\rho_0$). 

In this article, we review selected investigations of the high-density behaviour of the EoS derived from heavy-ion collision experiments 
with incident energies ranging from several hundred MeV up to approximately 2 GeV per nucleon. 
We also summarise constraints on the EoS of isospin-symmetric matter obtained from analyses of kaon and pion production and from measurements of collective flow observables.
For the symmetry energy, we discuss results obtained from studies of charged pion ratios and ratios of neutron-to-charged-particle elliptic flows. 
Furthermore, we describe how connections can be established between microscopic collisions (HIC), and macroscopic collisions like neutron-star mergers. 
In this context, estimates of neutron-star radii derived from heavy-ion collision results presented in this review are compared with astrophysical observations, 
including recent measurements enabled by gravitational-wave detections and X-ray satellite observations.
A multi-source analysis combining theoretical calculations, astrophysical data, and heavy-ion collision results to constrain neutron-star radii is presented 
as an example of the important contribution of heavy-ion studies in the era of multi-messenger astronomy. 
Finally, we outline several future prospects and challenges from both experimental and theoretical perspectives.

\section{Bridging heavy-ion collisions and neutron star mergers}\label{sec:bridge}

\subsection{The equation-of-state of nuclear matter}\label{sec:bridge1}

The EoS describes the relation between binding energy $E$, density $\rho$, pressure $P$, temperature $T$ and
isospin asymmetry $\delta=(\rho_n-\rho_p)/\rho$ of infinite nuclear matter, where $\rho_n$ and $\rho_p$ are the
neutron and proton densities, respectively \cite{BALi2008,Burgio2020}. It is conventionally split into an isospin symmetric matter part, independent of $\delta$, and an isospin dependent term, expressed
as a product of the symmetry energy (called also more academically "asymmetry energy") $E_{sym}(\rho)$, and the square of the isospin asymmetry parameter. 
For the case of $T=0$:
\begin{equation}
E(\rho,\delta)=E_{SNM}(\rho,\delta=0)+\delta^2E_{sym}(\rho)+o(\delta^4)
\end{equation}
The symmetry energy quantifies the energy cost of converting protons into neutrons in nuclear matter.

In thermodynamics, one derives the pressure from the energy per nucleon $E/A$ the following way, at constant temperature:    
\begin{equation}
P=\rho^2\frac{\partial E/A}{\partial \rho} \biggr\rvert_{T = const}
\end{equation}

In the field of the nuclear EoS, few key physical quantities are introduced to quantify the stiffness of the EoS. First, for SNM, the so-called incompressibility modulus:
\begin{equation}
K_0= 9 \rho^2  \frac{\partial^2 E/A}{\partial^2 \rho} \biggr\rvert_{\rho = \rho_0}
\end{equation}

Second, for the symmetry energy, the slope of its density dependence at saturation density:
\begin{equation}
L=3\rho_0 \frac{\partial E_{sym}}{\partial \rho} \biggr\rvert_{\rho = \rho_0}
\label{eq:L}
\end{equation}

\subsection{Temperatures and densities}\label{sec:bridge2}

\begin{figure}[!htb]
\centering
\includegraphics[width=0.99\textwidth]{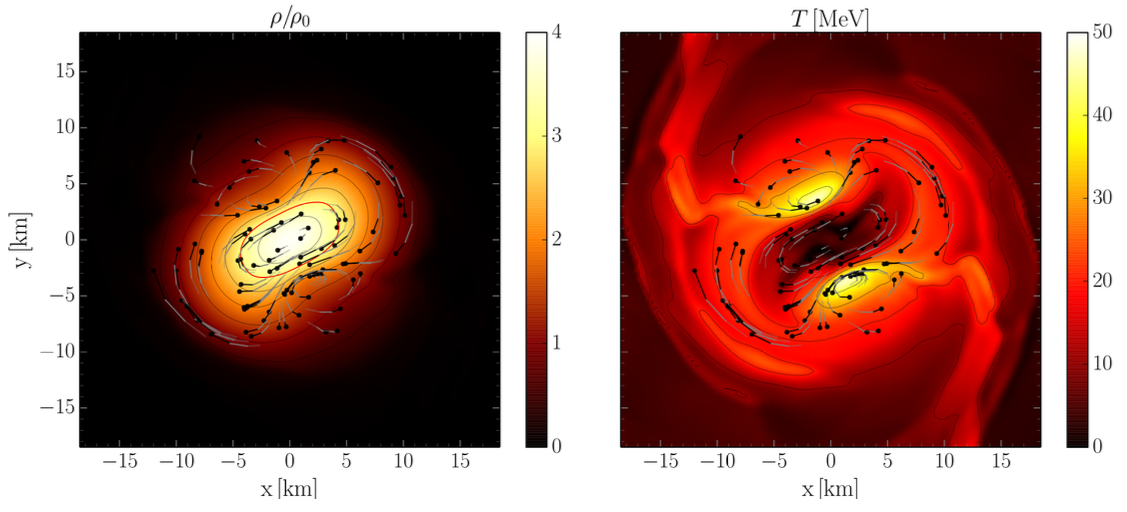}
\includegraphics[width=0.99\textwidth]{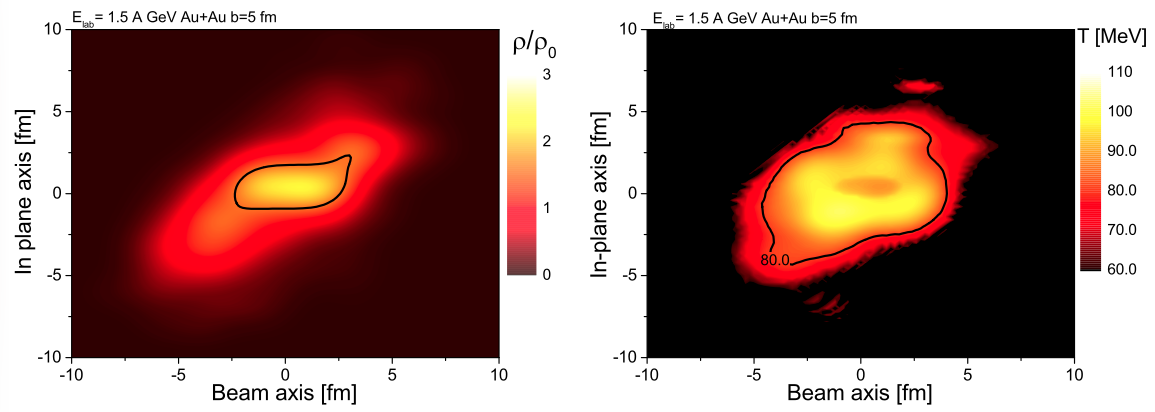}
\caption{Illustrations taken from \cite{Hanauske2017}. Simulations of a binary neutron star merger at a post-merger time of $t= 6.34 ms$ for the LS220-M132 binary (top panels) and of a gold on gold semi-central ($b=5fm$) collision at 1.5A~GeV incident energy at $t=15 fm/c$ (bottom panels). 
Right and left panels show respectively the temperature and density (relative to the nuclear saturation density) distributions in the rotation plane of both objects.
}\label{fig:Hanauske}
\end{figure}

According to transport models simulating HIC and numerical-relativity simulations of merging neutron star binaries \cite{Hanauske2017}, 
temperature and density created in the fireball (also called "participant") in HIC at intermediate energies (few hundred to few GeV per nucleon incident energy) 
mimic the density and temperature conditions created in binary neutron star mergers, as illustrated in Figure~\ref{fig:Hanauske}. Therefore, conditions of the EoS explored by
both microscopic and macroscopic collisions are similar. In both cases, temperatures remain quite moderate in comparison with HIC at ultra-relativistic energies.  

Since neutron stars in equilibrium are essentially cold (i.e. close to zero temperature), constraining their EoS using HIC requires the extraction of the cold nuclear EoS. 
This, in turn, implies that the temperatures reached in such collisions should not exceed the regime where the interplay between thermal effects 
and the zero-temperature EoS becomes too complex. At moderate temperatures, transport models used to interpret heavy-ion data can 
still account for thermal contributions through the momentum dependence of the mean-field potentials, 
thereby allowing the underlying non-thermal (cold) component of the EoS to be constrained.

\subsection{Relation between squeeze-out in HIC and tidal deformability of neutron stars}\label{sec:bridge3}

Other correspondencies between neutrons stars and nuclei, despite the vast difference in scale, can be found on the one side between 
the tidal polarizability (or deformability) of neutron stars and the squeeze-out amplitude of the fireball in HIC. 
Both are ruled on the first order by the stiffness of the EoS  \cite{Hinderer2010,ALF2018}.  
The analysis of the detected gravitational waves (GW) generated by the collision of neutrons stars made it possible to determine the tidal deformability, $\Lambda$. 
This quantity characterizes how strongly a star’s mass distribution is deformed into a quadrupole shape by the gravitational influence 
of its companion and therefore depends on the EoS. 
Tidal effects leave measurable imprints on the gravitational-wave signal during the inspiral phase of binary neutron stars, enabling constraints on the EoS.
In HIC at intermediate energies, in semi-central collisions, the participant region is highly compressed and flows afterwards preferentially perpendicular to the reaction plane, creating the so-called "negative elliptic flow" which shows the strongest sensitivity to the stiffness of the nuclear EoS.

\subsection{Relation between neutron skin of nuclei and neutron star radius}\label{sec:bridge4}

On the other side, neutron star radii find a correspondence with the thickness of the neutron skin of neutron-rich heavy-nuclei. 
It arises because both observables are governed by the same part of the nuclear EoS — specifically the density dependence of the nuclear symmetry energy.
In neutron-rich nuclei (e.g., $^{208}Pb$), the number of neutrons exceeds the number of protons. As a result, due to the force balance, neutrons extend further out than protons, forming a neutron skin. 
The neutron-skin thickness is defined as $\Delta R_{np}=R_n-R_p$ where $R_n$ and $R_p$ are respectively the root-mean-square radii of the neutron and proton distributions.
The thickness of this skin is controlled by the symmetry energy. If the symmetry energy increases rapidly with density (a “stiff” symmetry energy), 
neutrons experience stronger pressure and tend to spread out more, leading to a larger neutron skin.
Neutron stars consist primarily of neutron-rich matter at densities ranging from very low densities (in the atmosphere and crust) up to several times that value (in the core).
The pressure of neutron-rich matter near nuclear saturation density determines the size of neutron stars (their radii) and the extent of neutron skins in heavy nuclei.
A stiffer symmetry energy produces a larger pressure in neutron-rich matter, a thicker neutron skin in nuclei, and larger neutron-star radii.
Thus, measurements of neutron-skin thickness in heavy nuclei provide laboratory constraints on the EoS that also governs neutron-star structure around saturation density.

\subsection{A large variety of experimental observables for constraining the EoS over a broad range of density}\label{sec:bridge5}

\begin{figure}[!htb]
\centering
\includegraphics[width=0.95\textwidth]{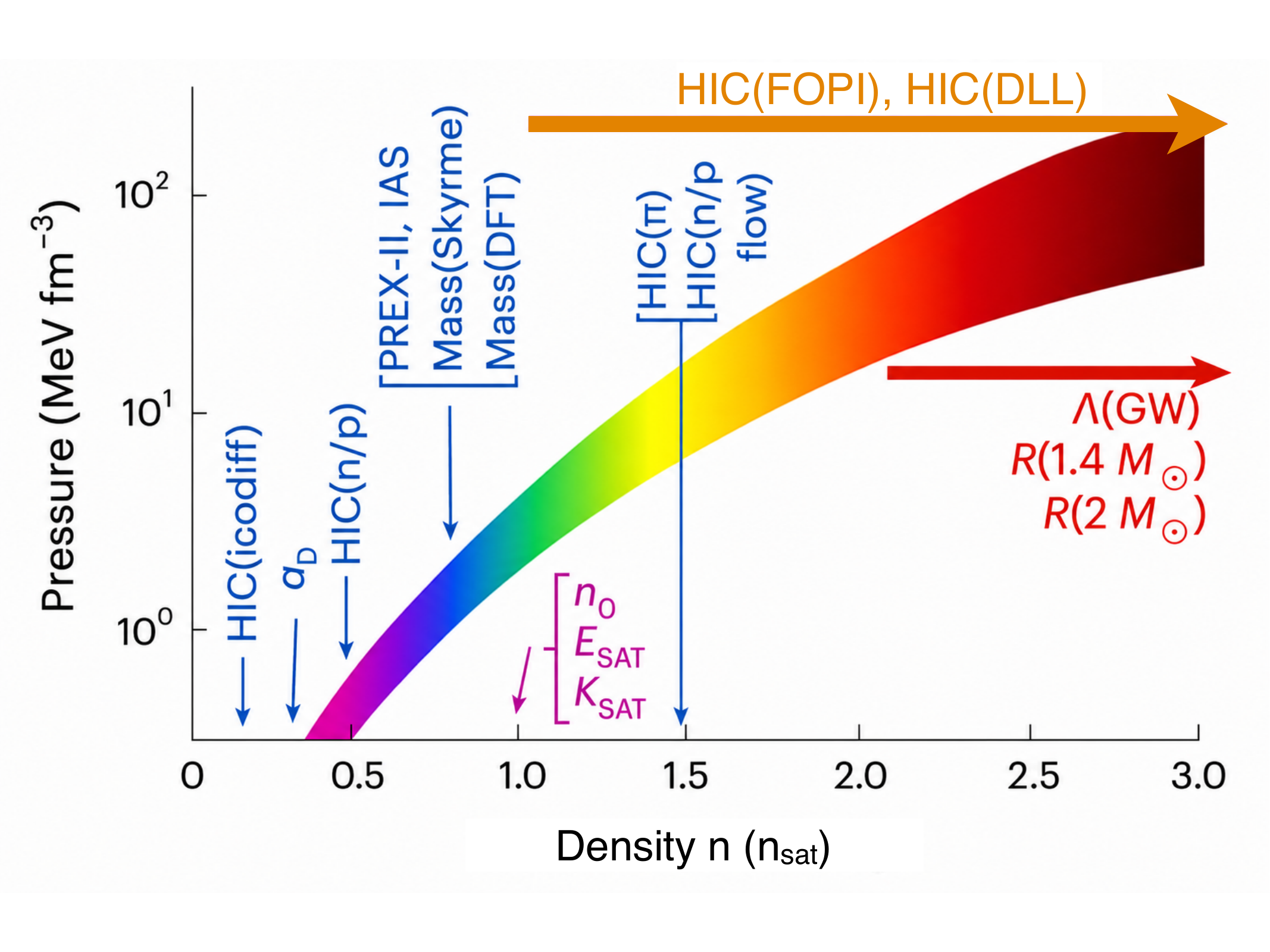}
\caption{Adapted from \cite{Tsang2024}. Constraints from nuclear experiments and astronomy observations
with their corresponding sensitive densities.
}\label{fig:TsangNature}
\end{figure}

There exists a large variety of experimental observables that can constrain the EoS, with their own density domain of sensitivity.
This is illustrated by Figure~\ref{fig:TsangNature}. There, the red horizontal arrow indicates the high-density region (above twice saturation density) probed by neutron star observations (GW, kilonovae and X-ray observations). The orange horizontal arrow indicates the domain of sensitivity that has been so far reached by HIC studies. 
The magenta-blue area concerns observables constraining the sub-saturation density domain.  

It is important to note that the average density in a neutron star is about $2n_{sat}$. Therefore, apart from the crust, that is sensitive to the low density properties of the EoS,  bulk properties of neutron stars are ruled by the high-density sector of the EoS. Therefore, HIC probes like flow, particle (pions, kaons, etc.) production are the most meaningful to complement informations delivered by multi-messenger astronomy.

Table \ref{tab:methods} details which component of the EoS (SNM or $E_{sym}$) can be probed by each nuclear physics observable.

\begin{table}[h]
\caption{Components of the nuclear EoS and density interval that are probed (indicated by the $\surd$ symbol; in brackets means ‘to be confirmed’) by various experimental probes of nuclear experiments}\label{tab:methods}%
\begin{tabular}{@{}llll@{}}
\toprule
Observable & SNM  & $E_{sym} $ & density interval of sensitivity ($n_{sat}$ units)\\
\midrule
Isospin diffusion (HIC)\footnotemark[1]&    &  $\surd$  & $0.5-1$  \\
Electric dipole polarisability $\alpha_D$ (GDR) \footnotemark[2] &    & $\surd$  & $\approx 0.5$  \\
n/p yield ratio (HIC) \footnotemark[3]  &  &  $\surd$ & $0.5-0.7$  \\
Neutron skin thickness (PREX)\footnotemark[4]  &  &  $\surd$ & $\approx 0.7$  \\
Isobaric analog states (IAS)\footnotemark[5]  &  &  $\surd$ & $0.5-0.7$  \\
Nuclear masses\footnotemark[6]  &  &  $\surd$ & $\approx 0.7$  \\
Isovector giant quadrupole resonance\footnotemark[7]  &  &  $\surd$ & $\approx 0.7$  \\
Isoscalar giant monopole resonance\footnotemark[8]  & $\surd$  &   & $0.7-1$  \\
Transversal expansion (HIC)\footnotemark[9]  & $\surd$  & ($\surd$)  & $>1$  \\
Pion yields (HIC)\footnotemark[10]  &  & $\surd$  & $>1.5$  \\
Kaon yields (HIC)\footnotemark[11]  & $\surd$ & ($\surd$)  & $>2$  \\
Flow (n, p, light clusters) \footnotemark[12]  &  $\surd$ & $\surd$  & $>2$  \\

\botrule
\end{tabular}
\footnotetext{References:}
\footnotetext[1]{\cite{Tsang2004,Tsang2009}}
\footnotetext[2]{In giant dipole resonances \cite{Tamii2011,Piekarewicz2012}}
\footnotetext[3]{\cite{Morfouace2019}}
\footnotetext[4]{\cite{PREX2021,RocaMaza2011}}
\footnotetext[5]{\cite{Danielewicz2014}}
\footnotetext[6]{Using Skyrme interaction or density functional theory (DFT), double magic nuclei \cite{Brown2013} and isotope binding
energy differences \cite{Zhang2013}}
\footnotetext[7]{\cite{RocaMaza2013}}
\footnotetext[8]{\cite{Blaizot1980,Youngblood1999,Khan2013}}
\footnotetext[9]{\cite{Stoecker1980}}
\footnotetext[10]{\cite{BALi2002,BALi2005,Estee2021}}
\footnotetext[11]{\cite{Aichelin1985,Hartnack2012,Sturm2001,Lopez2007}}
\footnotetext[12]{\cite{Danielewicz2002,ALF2016,ALF2018}}

\end{table}

\noindent

\section{Constraining the symmetric nuclear matter EoS at high density}\label{sec:SNM}

\subsection{Kaon production}\label{sec:SNM1}

KaoS at GSI Darmstadt has been a pioneering experiment constraining the isoscalar EoS by measuring kaon yields at sub-threshold energies,
where the sensitivity to the stiffness of the EoS is the largest.  

The production of positively charged kaons in HIC at sub-threshold energies ($E_{thr NN}=1.6$ GeV) proceeds through multistep processes 
in which intermediate $\Delta$ resonances and pions act as energy reservoirs. Consequently, the production rate depends strongly on the density reached during the collision. 
As the maximum compression increases, the number of secondary interactions rises, enhancing the probability of $K^+$ production.
In contrast, the threshold energy for $K^-$ production in nucleon–nucleon collisions is significantly higher ($E_{thr NN}=2.1$~GeV) due to its different quark content, 
$(\bar{u}s)$, compared with $(u\bar{s})$ for $K^+$. At sub-threshold energies, $K^-$ mesons are mainly produced via pion–hyperon reactions. The hyperons ($Y$)
 are generated together with $K^+$ in strangeness-conserving processes such as $\pi + N \rightarrow Y + K^+$, implying that $K^-$ production is closely linked to 
$K^+$ production. However, while $K^-$  mesons undergo strong absorption in nuclear matter, 
$K^+$  interact only weakly with the surrounding medium and can leave the reaction zone largely undisturbed. Therefore, 
$K^+$ mesons provide a sensitive probe of the maximum density achieved in HICs, as also supported by microscopic transport calculations \cite{Aichelin1985}. 
It should be noted that several reaction cross sections required as input for transport simulations are not experimentally accessible 
and are partly derived from theoretical models.

\begin{figure}[!htb]
\centering
\includegraphics[width=0.7\textwidth]{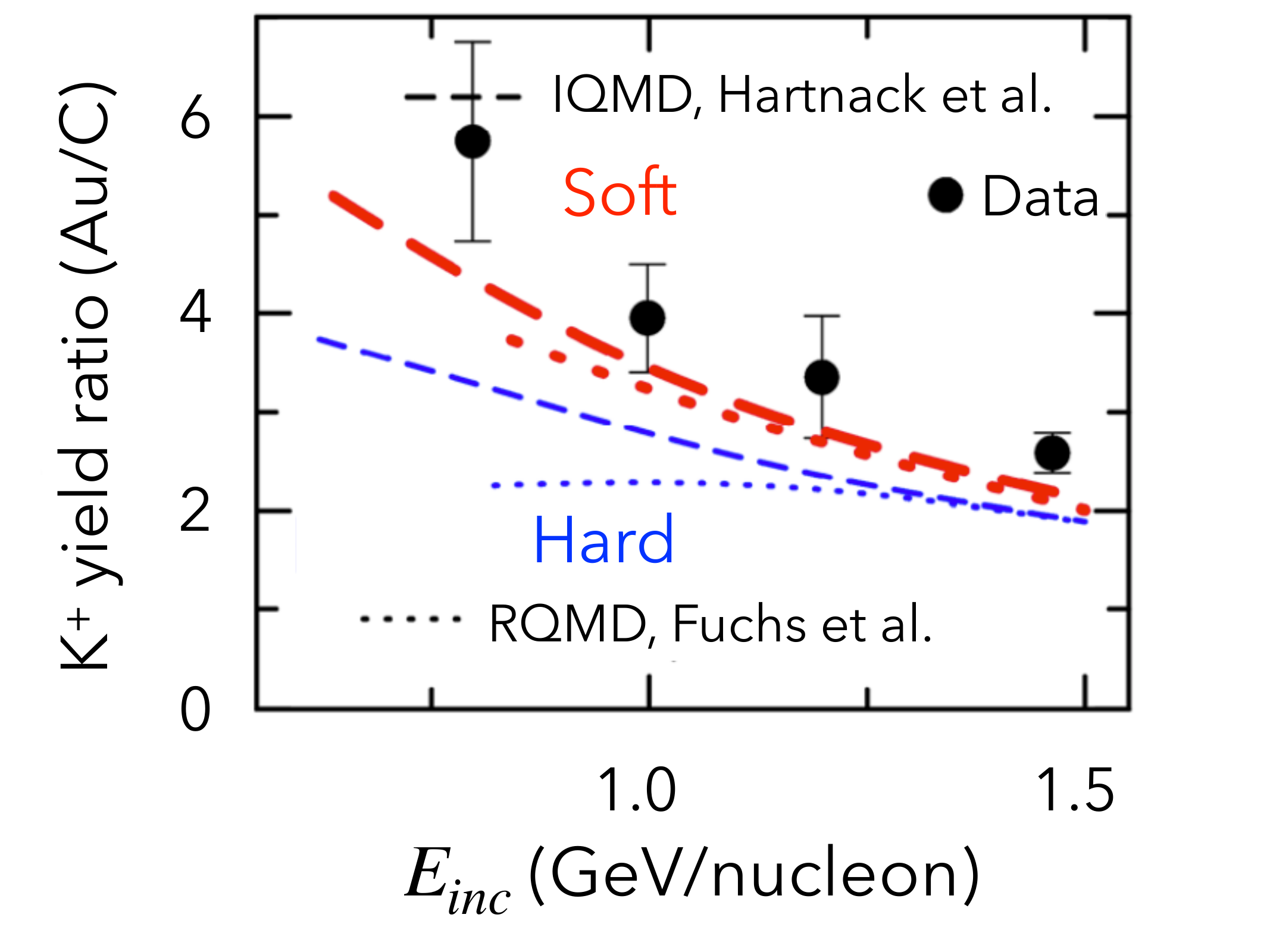}
\caption{Adapted from \cite{Hartnack2012}. 
KaoS experimental data for the ratio of $K^+$ production in \AuAu collisions to that in C~+~C collisions as a function of the incident energy are shown as black points. 
Predictions obtained with the IQMD model are represented by long dashed lines, while those from the RQMD model are indicated by short dashed lines. 
Calculations performed using a soft (hard) equation of state are displayed in red (blue). 
}\label{fig:KaoS}
\end{figure}

The KaoS collaboration measured $K^+$ production in \AuAu collisions at beam energies between 600 MeV/nucleon and 1.5 GeV/nucleon. 
Measurements in C+C collisions were used as a reference system to reduce the sensitivity to uncertainties in the elementary cross sections 
when comparing with theoretical predictions. The ratio of $K^+$ yields in \AuAu and C~+~C collisions is shown in Figure~\ref{fig:KaoS} 
and compared with calculations with the transport models IQMD \cite{Hartnack1998} and RQMD \cite{Fuchs2002}. 
In both cases, the data are best reproduced when a soft nuclear EoS is employed. 
Moreover, this conclusion remains robust against variations of the elementary cross sections, the inclusion of in-medium kaon effects, 
and changes in the momentum dependence of the nucleon–nucleon interaction in the transport models \cite{Hartnack2006}.

\subsection{Flow}\label{sec:SNM2}

In HIC at relativistic energies, the overlap region of the nuclei forms a hot and dense fireball of hadronic matter 
that expands and emits baryons and mesons toward the detectors.
The phase-space distribution of emitted particles is strongly influenced by the compression achieved in the collision zone 
and is therefore sensitive to the EoS of dense nuclear matter. 
A key observable used to probe the EoS is the flow. 
It is defined from azimuthal distribution of emitted particles which can be expanded as the following Fourier decomposition:

\begin{equation}
\frac{dN}{d(\phi-\Psi_{RP})} \propto 1 + 2 \sum_{n=1}^{\infty}v_n cos[n(\phi-\Psi_{RP})]
\label{eq:flow}
\end{equation}

where $\phi$ is the azimuthal angle of the particle, $\Psi_{RP}$ is the reaction-plane angle, $v_n$ are the flow coefficients.

Experimentally, the reaction plane is reconstructed event by event from the azimuthal distribution of detected particles, 
and the flow coefficients are corrected for finite resolution effects \cite{Andronic2006}. 

The amplitude of the directed flow is the first coefficient, which can be calculated this way: 
\begin{equation}
v_1 = \langle \cos\big(\phi-\Psi_{RP}\big) \rangle
\label{eq:v1}
\end{equation}
Positive (negative) $v_1$ indicates that particles are preferentially emitted in the direction (opposite side) of the projectile side.

The amplitude of the elliptic flow $v_2$, is defined as the second coefficient and can be calculated this way:
\begin{equation}
v_2 = \langle \cos2\big(\phi-\Psi_{RP}\big) \rangle
\label{eq:v2}
\end{equation}
 
Positive $v_2$ indicates preferential in-plane emission, while negative $v_2$ corresponds to out-of-plane emission.
Transport-model calculations have shown that the negative elliptic flow of protons and light clusters at mid-rapidity in collisions 
with incident energies between about 100 MeV/nucleon and 2 GeV/nucleon is particularly sensitive to the nuclear EoS \cite{Danielewicz2002,ALF2016,ALF2018,Reisdorf2012}.
This dependence is predicted by both quantum molecular dynamics (QMD) and Boltzmann–Uehling–Uhlenbeck models
 \cite{Danielewicz2002,Hartnack1998,ALF2016,Reisdorf2012,Wang2018,Cozma2024}. 
The origin of this out-of-plane emission is attributed to the interaction between the expanding fireball and the spectator matter, which shadows in-plane emission 
and enhances particle emission perpendicular to the reaction plane \cite{ALF2018}. The resulting anisotropy depends strongly on the density gradients generated 
during the collision and therefore on the EoS. 
At higher beam energies (1–10 GeV/nucleon), the directed flow $v_1$ becomes similarly sensitive to the stiffness of the EoS \cite{Danielewicz2002}. 
Analyses of flow data from \AuAu collisions at the Bevalac and AGS accelerators have constrained the pressure of symmetric nuclear matter at densities up to 
$\approx 3-4n_{sat}$, although the interpretation becomes more uncertain due to increased resonance and meson production at these energies. 

Overall, HIC flow measurements at energies from a few hundred MeV/nucleon to about 10 GeV/nucleon favour a relatively soft EoS with incompressibility $K_0\lesssim 260$~MeV
when momentum-dependent interactions are included.
More precise constraints were obtained from elliptic-flow measurements by the FOPI Collaboration in \AuAu collisions between 0.4 and 1.5 GeV/nucleon. 
Comparison of proton and light-cluster flow data with simulations using the isospin-QMD (IQMD) model yielded an incompressibility $K_0=190\pm30$~MeV \cite{ALF2016},
 clearly favouring a soft momentum-dependent EoS over a stiff one. 
Later analyses using the UrQMD \cite{Li2005} and dcQMD \cite{Cozma:2014yna} models obtained consistent values, respectively 
$K_0=220\pm40$~MeV \cite{Wang2018} and $K_0=190^{+9}_{-11}$~MeV \cite{Cozma2024}. 
In particular, the work of Ref.~\cite{Cozma2024} provides improvements as it is
a comprehensive data analysis optimising various parameters and assumptions of the transport model, 
such as to narrow down constraints, and to better determine uncertainties. 
Recently, an analysis of FOPI flow data with the BUU-like model SMASH \cite{Tarasovicova2024} has concluded with a similar soft EoS, 
with the difference that the higher-energy data favour a stiffer EoS according to this model, which may indicate a stiffening of the EoS at $\approx 3n_{sat}$.   

\begin{figure}[!htb]
\centering
\includegraphics[width=0.6\textwidth]{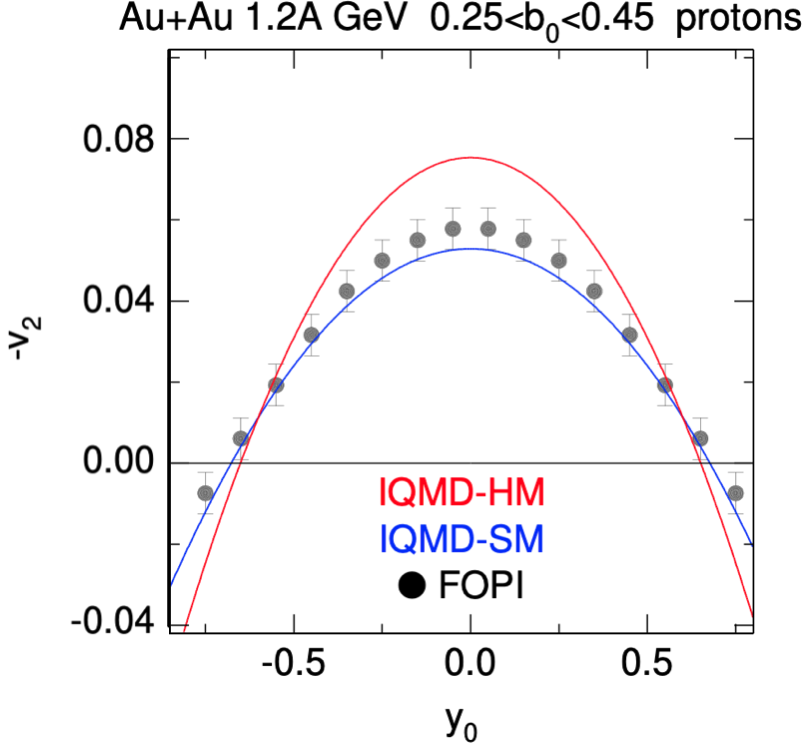}
\caption{Reprinted from Ref.~\cite{ALF2016}. 
Proton elliptic flow data from \AuAu mid-central collisions at 1.2 GeV/nucleon incident energy measured by the FOPI set-up, as a function of the rapidity 
scaled to that of the projectile (black dots), and IQMD-SM/HM simulations (blue/red curves, respectively).
See text and source in Ref.~\cite{ALF2016} for further explanations.
}\label{fig:FOPI1}
\end{figure}

As an illustration of the sensitivity of $v_2$ to $K_0$, Figure~\ref{fig:FOPI1}, taken from Ref.~\cite{ALF2016}, shows proton elliptic-flow data ($-v_2$)
from the FOPI Collaboration (black points with error bars) together with simulations from the IQMD transport model \cite{Hartnack1998}. 
The calculations are performed using a stiff (HM, red) and a soft (SM, blue) EoS, including a momentum-dependent interaction which was required to reproduce other observables (see details below). 
The results are presented as a function of the rapidity in the centre of the system scaled to that of the projectile $y_0=y/y_{proj}$
for \AuAu collisions at an incident energy of 1.2 GeV/nucleon.
Additional data covering incident energies from 0.15 to 1.5 GeV/nucleon and various system sizes and centralities are presented in Ref.~\cite{ALF2016}. 
Plotting $-v_2$ highlights an enhancement around mid-rapidity, corresponding to predominantly out-of-plane emission. This effect, originally termed “squeeze-out” 
in early theoretical work \cite{Stoecker1982}, is now generally referred to as elliptic flow, reflecting the combined influence of spectator shadowing and density gradients
as demonstrated in Ref.~\cite{ALF2018}. At larger rapidities $|y_0|$, the flow changes sign, indicating a transition to predominantly in-plane emission.

An important finding of this study is that in order to improve the accuracy in constraining $K_0$ from the elliptic flow, one must include the rapidity dependence of $v_2$, consider various bombarding energies, and include light clusters whose $v_2$ turns out to be more sensitive to $K_0$ than protons alone. 

Overall, constraints from the elliptic flow are compatible with those derived from sub-threshold kaon production measurements mentioned in Section~\ref{sec:SNM1}.

An important feature of flow studies, as they are done on the basis of transport models simulations of collisions,
 is the compulsory inclusion of momentum dependent interactions (mdi) in those models, which are sometimes erroneously neglected in the literature. 
Adding a mdi is a natural way to account for the thermal part of the dynamics adding to the mean-field potential which corresponds to a cold EoS.  
This decoupling between cold EoS and thermal motion remains valid as long as temperatures reached in HIC do not exceed few tens of MeV. 
There are historical evidences in the literature that the mdi is required in HIC at intermediate energies to reproduce some observables. 
The first one is the directed flow at large rapidity (close to the spectators) in $Ni+Ni$ semi-central collisions as shown in Ref.~\cite{Andronic2003}. 
The second evidence is that, according to Ref.~\cite{Hartnack2006}, kaon yields can only be explained with a soft EoS, independently of the inclusion of mdi. 
And flow data without mdi would always artificially advocate for a stiff EoS ($K_0>280$~MeV) as recently shown by \cite{Zhou2025}, in contradiction with kaon data.
Omitting the mdi is essentially equivalent to predicting a hot EoS, which is not comparable to a zero temperature EoS. 

\subsection{Probed densities}\label{sec:SNM3}

Deducing the most probable EoS from a given observable is not enough. 
The importance of identifying the density that a given observable actually probes has been early highlighted by \cite{Horowitz:2014bja, ALF2016, Russotto2016}, and recently by \cite{Lynch2022}.
This approach ensures that constraints on the nuclear equation of state (EoS) are restricted to the relevant density region, 
thereby avoiding unreliable extrapolations to other densities.

Therefore, the second key information to be provided is the probability distribution of density that this particular observable probes with the experimental data set used.
As we have seen in Section~\ref{sec:bridge5}, various observables do not have the same density range of sensitivity. In addition, it is expected that flow studies done 
on HIC performed at high incident energies will naturally tend to probe larger densities than at low incident energies as shown e.g., in Refs.~\cite{BALi2002-2,ALF2016}.    
Unfortunately, this good practice is often neglected in the literature. 
Several works like \cite{ALF2016,Russotto2016} propose methods to deduce this density profile.  

\begin{figure}[!htb]
\centering
\includegraphics[width=0.95\textwidth]{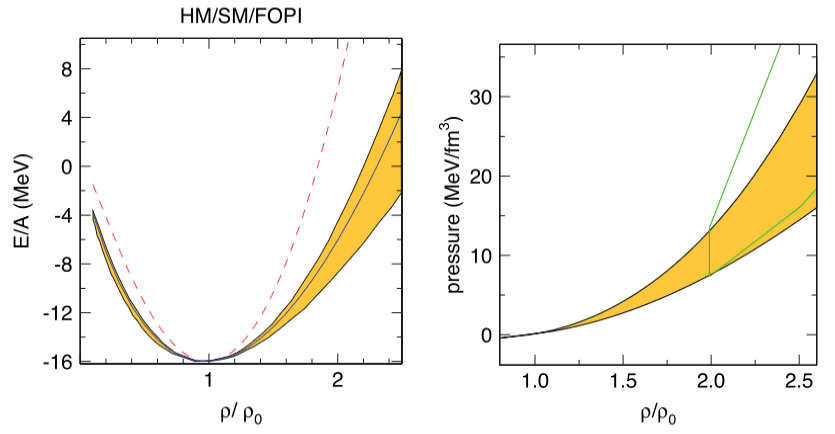}
\caption{(Reprinted from Ref. \cite{Russotto2023}) Left panel: The yellow band delimits the FOPI EoS constraints for SNM from \cite{ALF2016} corresponding to $K_0 = 190 \pm 30$~MeV, compared with a hard and soft EoS's (respectively dashed red and full blue curves).  
Right panel: The same nuclear EoS in terms of pressure versus density as obtained
from FOPI data (yellow band). The area framed with green lines originates from the AGS and Bevalac data analysed by \cite{Danielewicz2002}.
}\label{fig:FOPI2}
\end{figure}

These studies indicate that elliptic-flow data at intermediate energies probe densities between about 0.7 and 3$n_{sat}$. Figure~\ref{fig:FOPI2} summarises 
the EoS constraint obtained from FOPI data \cite{ALF2016}.

\begin{figure}[!htb]
\centering
\includegraphics[width=0.75\textwidth]{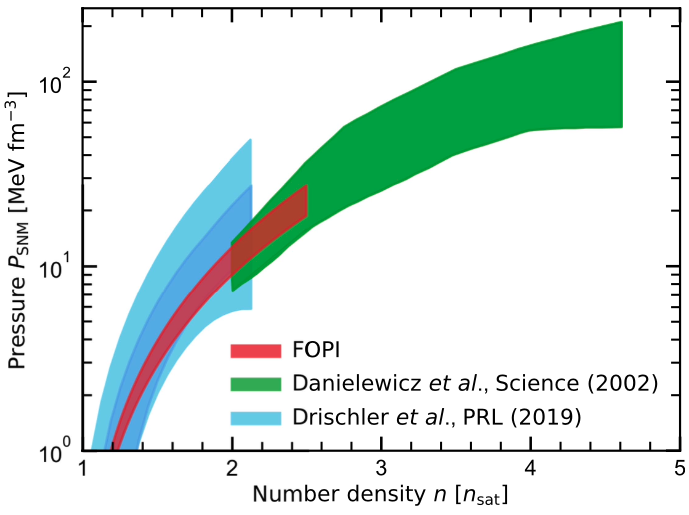}
\caption{Reprinted from Ref.~\cite{Huth2022}. 
Comparison of the pressure of SNM (in logarithmic scale) for experiment and theory. The pressure band from the FOPI
experiment (at GSI) at the $1\sigma$ level (red) for the incompressibility is consistent with the
chiral EFT constraint from \cite{Drischler2019} at N2LO (light blue) and N3LO (dark
blue). The constraint from AGS and Bevalac data analysed by \cite{Danielewicz2002} (green)
prolongates the FOPI expectation with no tension and a lesser accuracy.  
}
\label{fig:Huth1}
\end{figure}

Adding the constraints obtained by \cite{Danielewicz2002} from AGS and Bevalac flow data at higher energies, 
one can draw an EoS constraint over a large span of densities extending up to $\approx 5n_{sat}$ as illustrated by Figure~\ref{fig:Huth1}. 
There, the FOPI pressure curve was obtained by combining conclusions of \cite{ALF2016} (with $K_0 = 190 \pm 30$~MeV) and \cite{Wang2018} (with $K_0=220\pm40$~MeV),
being therefore slightly stiffer than what is shown in Figure~\ref{fig:FOPI2}.  

\section{Constraining the symmetry energy at high density}\label{sec:Esym}

\subsection{Flow}\label{sec:Esym1}

\begin{figure}[!htb]
\centering
\includegraphics[width=0.6\textwidth]{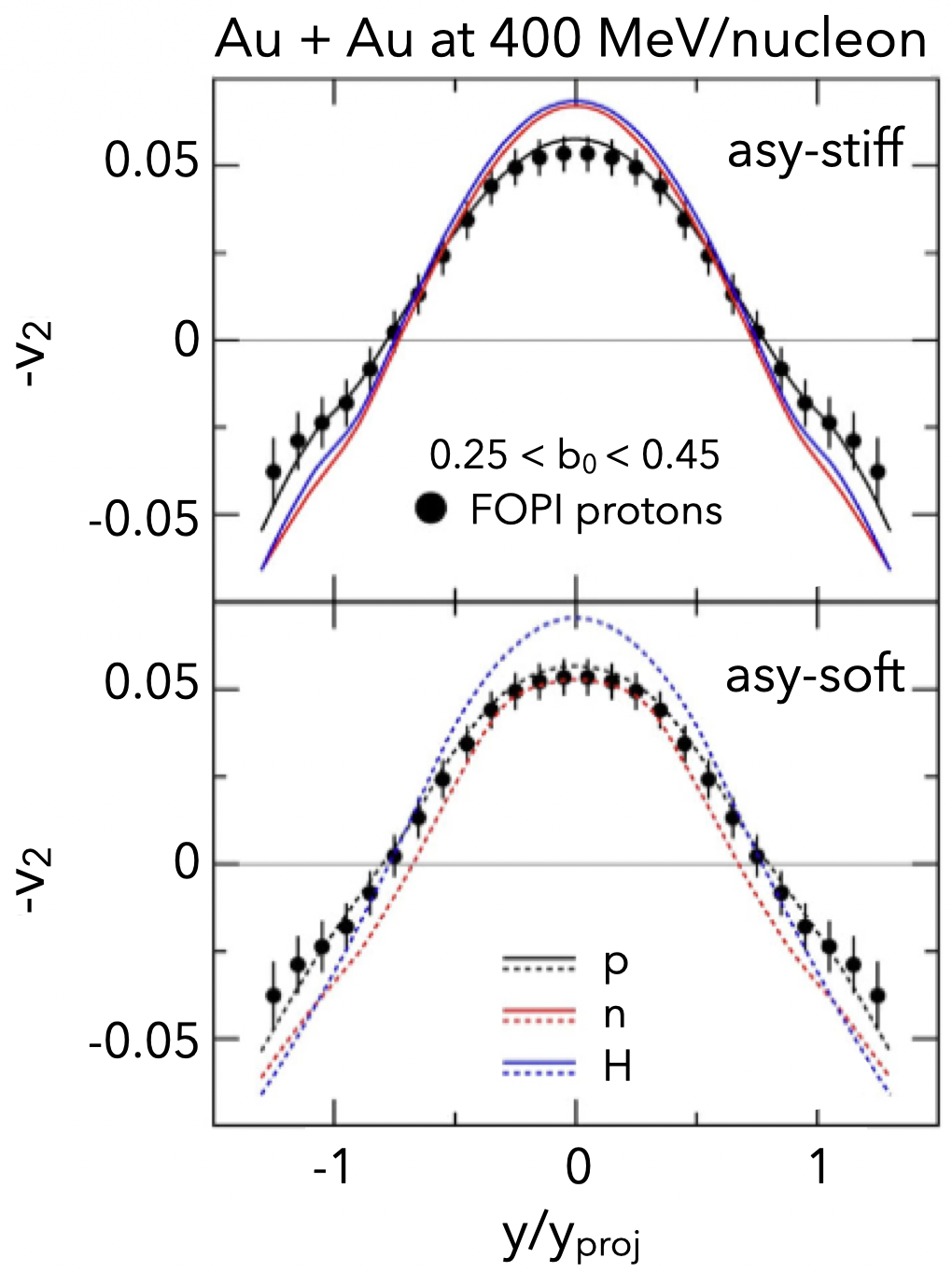}
\caption{(adapted from \cite{Russotto2023})
UrQMD (version from \cite{Li2005}) predictions for the elliptic flow in opposite sign ($-v_2$) of neutrons, protons, and hydrogen are shown as a function of the scaled rapidity 
$y_0$ for \AuAu collisions at 400 MeV/nucleon in the reduced impact parameter range $ 0.25 <b_{0}=b/b_{max} < 0.45$. 
The calculations are performed using parameterisations of the symmetry-energy density dependence corresponding to a stiff (top panel) and a soft (bottom panel) scenario. 
For comparison, the proton elliptic flow data measured by the FOPI Collaboration are also displayed in both panels (black solid circles).
}\label{fig:Esym1}
\end{figure}

The so far most accurate observable to constrain the symmetry at large densities is the elliptic flow of neutrons compared to that of charged particles, in particular protons. 
The first evidence for that was found by studying \AuAu collisions at 400 MeV/nucleon simulated using the UrQMD transport model in \cite{Li2005}.
We present in Figure~\ref{fig:Esym1} how the strength of the symmetry energy influences differently the elliptic flow of neutrons and protons 
(and more generally charged particles). 

\begin{figure}[!htb]
\centering
\includegraphics[width=0.47\textwidth]{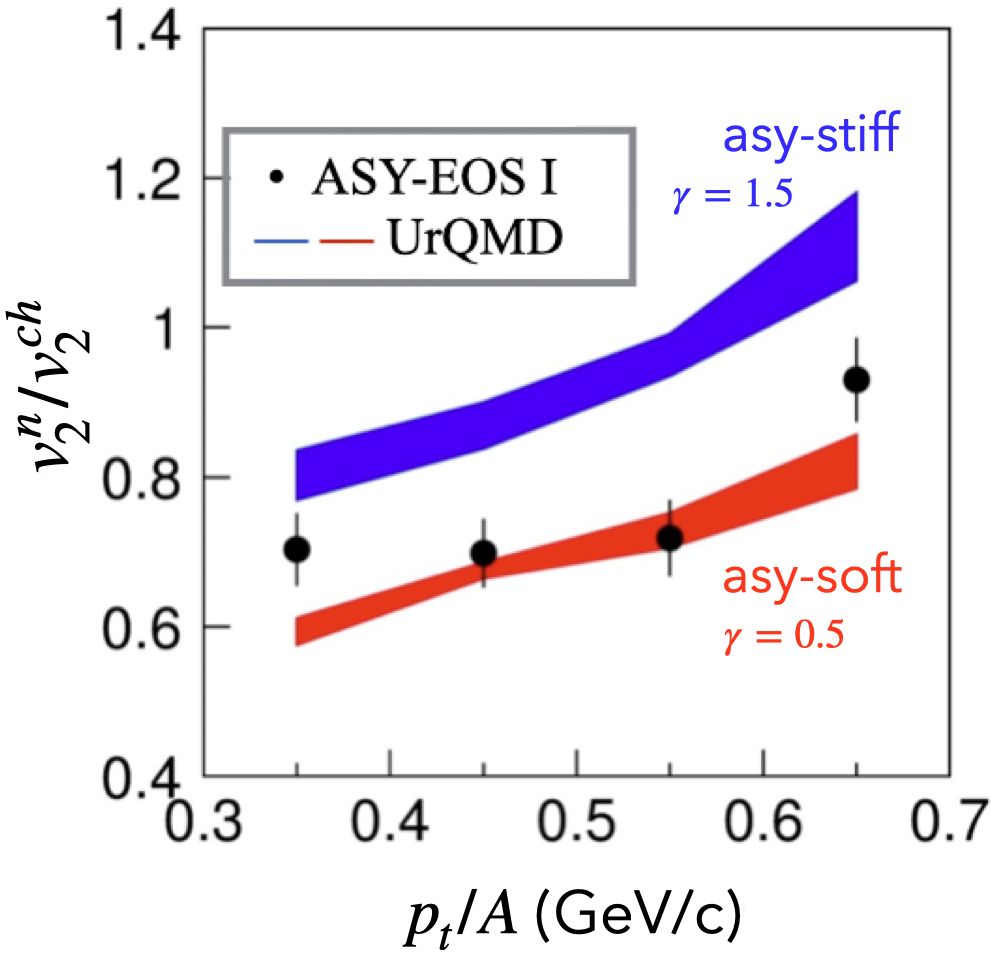}
\includegraphics[width=0.47\textwidth]{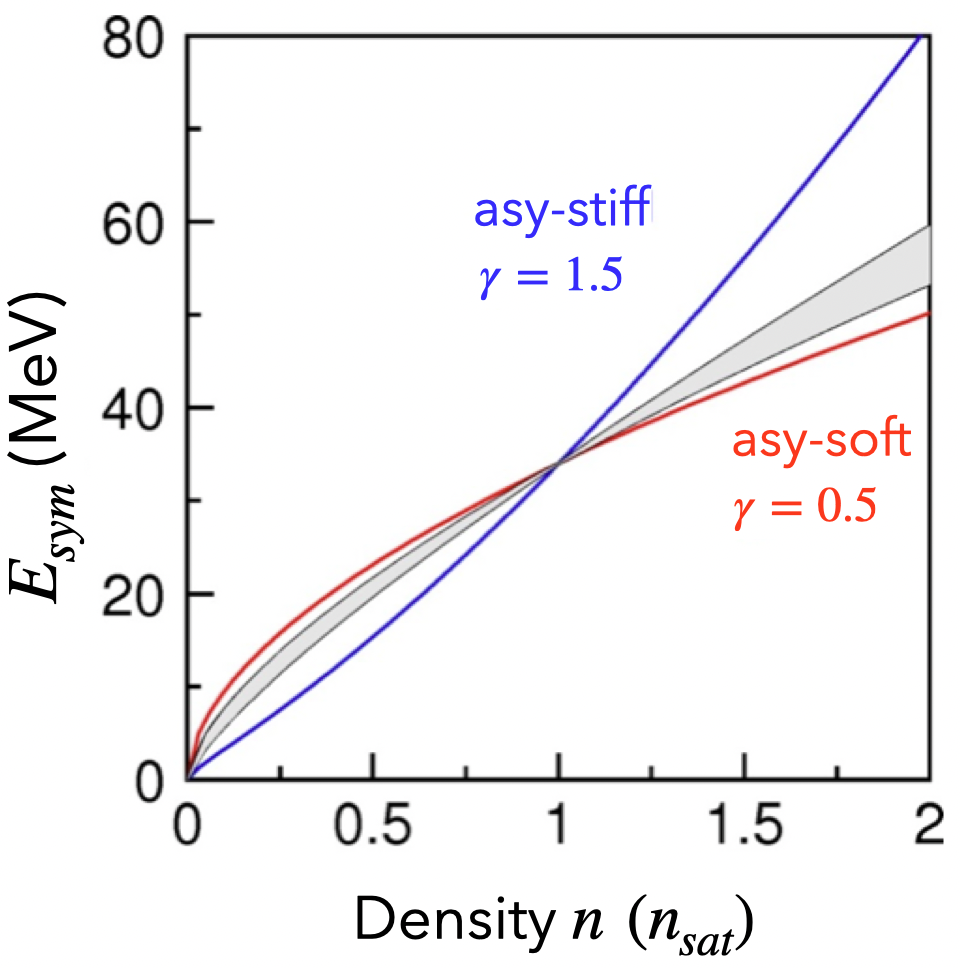}
\caption{(adapted from Ref.~\cite{Russotto2016})
Left panel: Elliptic flow ratio of neutrons over charged
particles in the same acceptance range for semi-central ($b<7.5 fm$) \AuAu collisions at
400 MeV/nucleon incident energy as a function of transverse momentum, $p_t/A$. The black markers represent the ASY-EOS
experimental data (first campaign). The blue and red bands represent expectations from UrQMD for respectively 
a stiff ("asy-stiff", $\gamma=1.5$) and a soft ("asy-soft", $\gamma=0.5$) density dependence of $E_{sym}$ as parametrised in this model
according to Equation~\ref{eq:potasy}. 
Right panel: Representation of the density dependence of $E_{sym}$ for the stiff (blue) and soft (red) parametrisation, and for the stiffness best reproducing the 
 experimental data (grey band). 
}\label{fig:Esym2}
\end{figure}

A stiff symmetry energy leads to a larger magnitude of neutron $v_2$ at mid-rapidity than that of protons, 
while the opposite but weaker behaviour occurs for a soft symmetry energy. This results in a characteristic inversion of neutron and proton elliptic flows. 
The effect originates from particles emitted perpendicularly to the reaction plane at mid-rapidity, which mainly stem from the expansion 
of the hot and dense fireball. For neutron-rich matter above saturation density, a stiff symmetry energy produces stronger repulsion for neutrons and 
stronger attraction for protons, enhancing neutron elliptic flow. 
For protons, however, the difference between stiff and soft cases is smaller because Coulomb and symmetry forces partially compensate each other.
To constrain the symmetry energy, it is therefore advantageous to use the ratio of neutron to proton elliptic flows rather than neutron flow alone, as mentioned early in \cite{Trautmann:2010at,Russotto2011}. 
Many transport-model ingredients 
-- such as the stiffness of the isoscalar EoS, in-medium nucleon–nucleon cross sections, nucleonic wave-packet width, and initialisation conditions -- 
affect neutrons and protons similarly and largely cancel in the neutron over proton $v_2$ ratio. These effects have been studied in detail in \cite{Cozma2011,Cozma2013}. 
The ratio also reduces experimental systematic effects, such as those related to reaction-plane resolution. Since the elliptic flow of $Z=1$ particles shows a behavior 
similar to that of protons, the comparison between neutron and hydrogen isotopes flows can also serve as a probe of the symmetry energy.
These conclusions, first obtained with the UrQMD model, have been confirmed by other QMD transport models such as IQMD and TüQMD \cite{Cozma:2014yna}, as well as by BUU-type approaches like the Stochastic Mean Field model of the Catania group \cite{Giordano2010}.
The first experimental constraint that has been obtained using the flow difference method has been presented in Ref.~\cite{Cozma2011} 
on the basis of FOPI-LAND data taken at GSI from \AuAu collisions at 400, 600 and 800 MeV/nucleon incident energy \cite{Leifels1993,Lambrecht1994}. 
The accuracy of the constraint has been further improved 
with the first ASY-EOS campaign taking place at GSI, which unlike the FOPI data, did not access the protons separately from other light charged particles. 
Results of this experiment presented in \cite{Russotto2016} are shown in Figure~\ref{fig:Esym2} where the transverse momentum $p_t$ dependence of the $v_2$ ratio 
between neutrons and light-charged particles (mostly $Z=1$), measured at mid-rapidity in \AuAu semi-central collisions at 400 MeV/nucleon incident energy,
 was used to constrain the density dependence of $E_{sym}$. 
   
To do so, various stiffnesses of $E_{sym}$ have been simulated by UrQMD. In this
transport code, the potential part ($E_{sym}^{pot}$) of the symmetry energy is parametrised by a power-law as a function of
the reduced density. It adds to the kinetic part ($E_{sym}^{kin}$) of it to build the total symmetry energy
\begin{equation}
E_{sym}(\rho)= E_{sym}^{pot}(\rho) + E_{sym}^{kin}(\rho) = 22~MeV\dot(\rho/\rho_0)^\gamma+12~MeV\dot(\rho/\rho_0)^{2/3}
\label{eq:potasy}
\end{equation}

The best agreement with the experimental data has been obtained using $\gamma=0.72\pm0.19$, systematics errors taken into account. 
The corresponding slope parameter L (see Equation~\ref{eq:L}) spans between $72 \pm 13$~MeV and $63 \pm 11$~MeV, depending on whether one takes the reasonable 
assumption that the symmetry energy at saturation density $S_0 = 34$ or $31$~MeV, respectively. 
A more recent and comprehensive analysis of FOPI flow and nuclear stopping data from HIC at similar incident energies by \cite{Cozma2024} with an upgraded version of the dcQMD transport model found an agreement by concluding
$L=63^{+10}_{-13}$~MeV, $S_0=35\pm1$~MeV. 

The sensitive density profile probed by the ASY-EOS experimental data has been found to span between 0.5 and 1.5 $n_{sat}$ at $1\sigma$.   

\begin{figure}[!htb]
\centering
\includegraphics[width=0.7\textwidth]{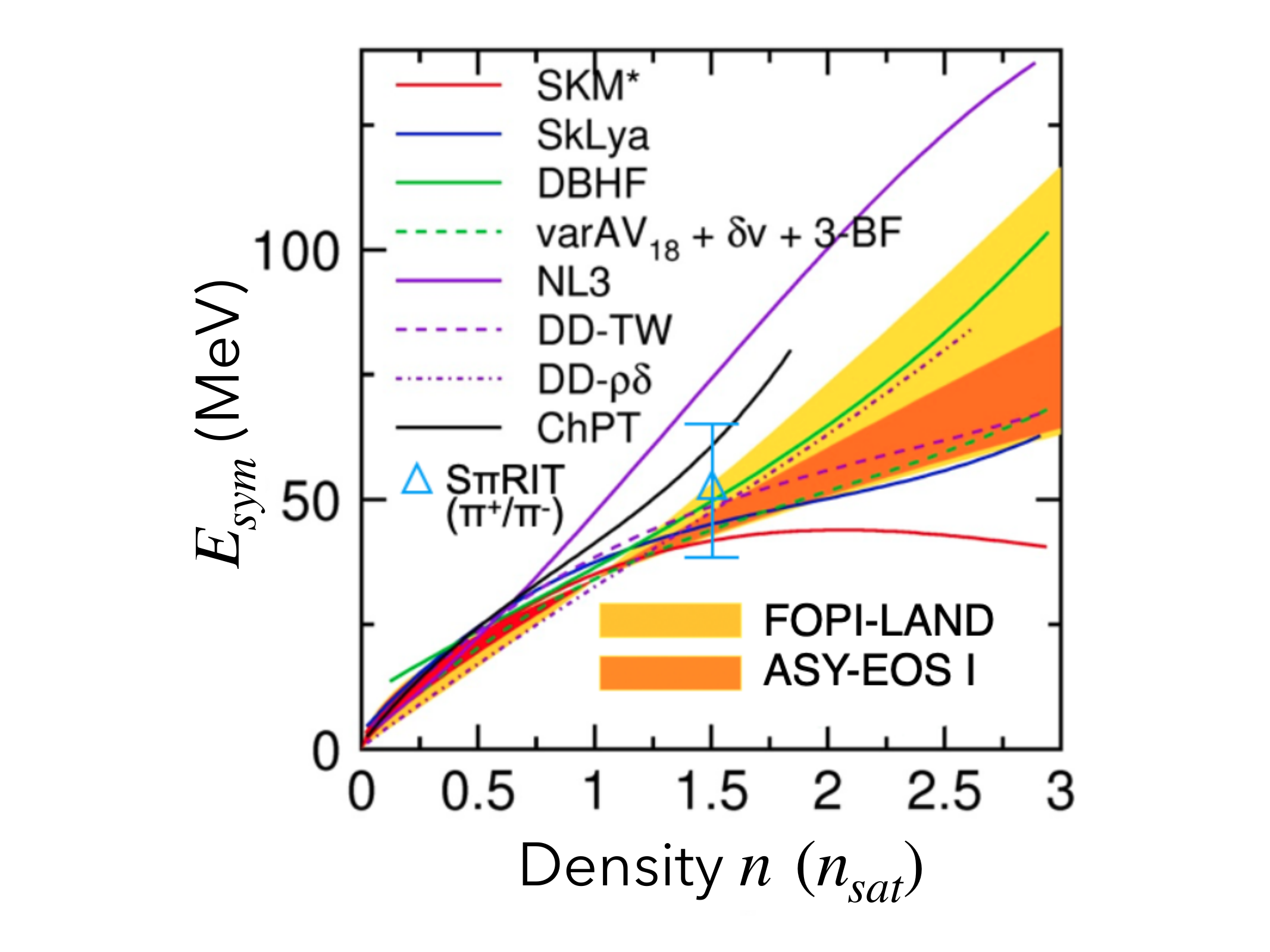}
\caption{(Adapted from \cite{Russotto2016})
Constraint deduced for the density dependence of the symmetry energy from ASY-EOS (first campaign, orange band) \cite{Russotto2016}
and FOPI-LAND (yellow band) \cite{Russotto2011} experiments, compared also to various ab-initio and phenomenological calculations \cite{Fuchs2006}, and to the constraint obtained by the S$\pi$RIT Collaboration from pion yield ratios (blue triangle) \cite{Estee2021}. Note that some rather old theoretical predictions do not represent anymore the state of the art as, for example, 
the result labeled ChPT which appears unrealistically stiff.
}\label{fig:Esym3}
\end{figure}

Figure~\ref{fig:Esym3} shows the outcome of the resulting ASY-EOS constraint of $E_{sym}(\rho)$ compared with the pioneering FOPI-LAND results.  
Both experiments agree and ASY-EOS exhibits a much improved accuracy. 

\subsection{Pion production}\label{sec:Esym2}

The ratio of charged pion multiplicities in HIC was proposed by \cite{BALi2002,BALi2005} as a probe of the density dependence of the nuclear symmetry energy. 
Soon after, measurements by the FOPI Collaboration for systems such as \AuAu became available 
and were used to extract constraints on the symmetry-energy slope parameter L \cite{Reisdorf2007}. 
However, analyses using different transport models produced widely varying results -- from very soft to very stiff symmetry energies -- 
revealing significant uncertainties in the modelling of pion production near threshold.
These discrepancies arise from incomplete knowledge of several ingredients affecting pion production, including in-medium pion–nucleon interactions, 
pion optical potentials, neutron-skin effects, and modifications of particle-production thresholds. 
Such effects can significantly influence pion multiplicities and their ratios, making consistent descriptions of pion and nucleon observables difficult.
To address model dependencies, an international collaboration of transport-model developers (TMEP for Transport Model Evaluation Project) 
performed systematic benchmark studies comparing different models and testing specific ingredients such as collision terms, Pauli blocking, 
and mean-field dynamics \cite{Xu2016,Zhang2018,Ono2019,Colonna2021,Wolter2022}. Despite significant progress, substantial differences among model predictions remain.

Recent experimental efforts have been made by the S$\pi$RIT Collaboration, measuring pion production in Sn~+~Sn collisions at 270 MeV/nucleon at RIKEN Radioactive Isotope Beam Factory.
Comparisons with several transport models show large variations in predicted pion multiplicities, although some models reproduce pion ratios reasonably well.
More robust constraints can be obtained from the high-transverse-momentum region of pion spectra, which is less sensitive to uncertain model ingredients. 
Using this approach, recent analyses with the dcQMD model presented in \cite{Estee2021} extracted a symmetry-energy slope parameter of $L \approx 80 \pm 38$~MeV
in association with a symmetry energy at saturation density $S_0 \approx 35 \pm 3$~MeV. Probed densities have been estimated to be about $1.5n_{sat}$. 
The resulting $E_{sym}(\rho)$ region of expectation is represented by the blue triangular marker in Figure~\ref{fig:Esym3}, 
showing a fair agreement with flow data of ASY-EOS and FOPI-LAND, but with a larger error bar.

In the continuation of the comprehensive analysis by M.D. Cozma in \cite{Cozma2024} based on flow and stopping data, the author recently included S$\pi$RIT, and FOPI pion yields obtained at higher incident energies. He demonstrated in \cite{Cozma2025NUSYM} that, due to larger uncertainties of the $\Delta$ potential, low (sub-threshold) beam energies probed by the S$\pi$RIT experiment are less favorable 
in terms of constraining power of the symmetry energy using pion yields. 
Therefore, this opens perspectives that both FOPI and HADES pion yields may improve the knowledge of $E_{sym}$ at high density with a fair accuracy.   

\subsection{Momentum dependence of the symmetry energy}\label{sec:Esym3}
Like for the isoscalar part (see Sect.~\ref{sec:SNM2}), the momentum dependence of the isovector part of the nuclear interaction — often expressed through the neutron/proton effective-mass splitting in asymmetric nuclear matter — is a key uncertainty of the EoS. It affects nucleon transport in neutron-rich systems, the density dependence of the symmetry energy, neutrino propagation and cooling in neutron stars, collective flows in heavy-ion collisions, and the composition and transport properties of dense astrophysical matter \cite{BALi2015}. Determining whether neutrons have a larger or smaller effective mass than protons in neutron-rich matter therefore has implications spanning finite nuclei, heavy-ion dynamics, and neutron-star structure.
Experimentally, constraints have mainly come from intermediate-energy heavy-ion collisions involving neutron-rich systems. Transport-model analyses have used observables sensitive to the momentum-dependent symmetry potential, such as neutron/proton spectral ratios, elliptic and directed flows, pion production, and isospin diffusion. Early studies, including the work of \cite{Coupland2016}, showed that double neutron/proton yield ratios and collective flows could discriminate between opposite signs of the mass splitting, although strong model dependencies remained. More recent efforts, such as \cite{Cozma2024}, combine improved detector capabilities, better neutron measurements, and systematic comparisons among transport codes to reduce uncertainties. Current evidence tends to favor $m_n^*>m_p^*$ in neutron-rich matter near and slightly above saturation density, but quantitative constraints remain incomplete because extracted results still depend sensitively on the treatment of in-medium cross sections, cluster production, and transport-model implementations.

\section{Confrontation of HIC and multi-messenger astronomy EoS constraints}\label{sec:comparison}

\subsection{A pioneering comparison}\label{sec:comparison1}

\begin{figure}[!htb]
\centering
\includegraphics[width=0.99\textwidth]{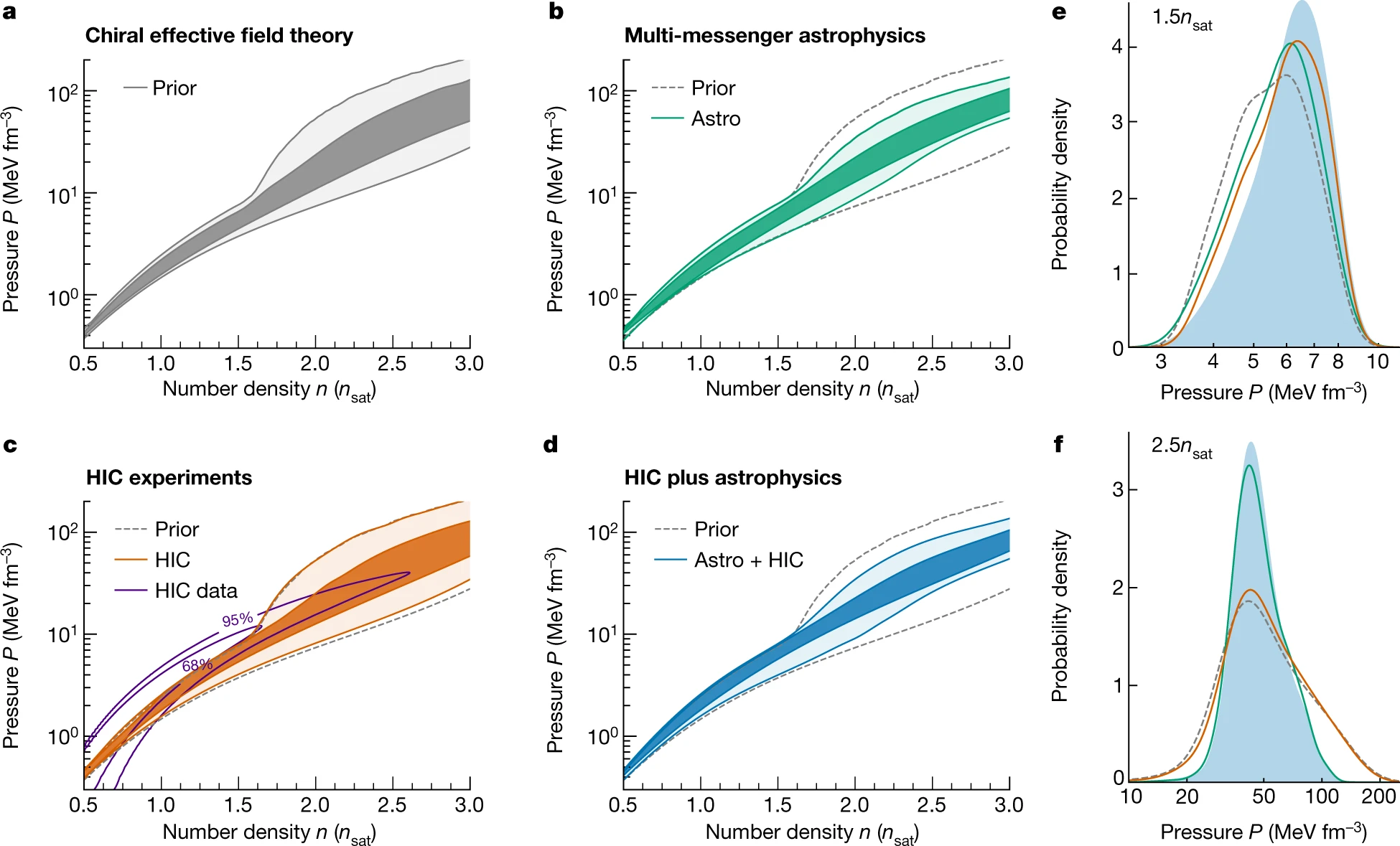}
\caption{(Reprinted from \cite{Huth2022}) Evolution of the pressure as a function of the baryon
number density for the EoS prior (a, gray), when including only data from multi-messenger neutron-star
observations (b, green), when including only HIC data (c, orange), and when combining both (d, blue).
The shading corresponds to the 95$\%$ and 68$\%$ credible intervals (lightest to darkest). The impact of the
HIC experimental constraint (HIC data, purple lines at 95$\%$ and 68$\%$) on the EoS is shown in panel c.
In panels (b) through (d), the 95$\%$ prior bound is also shown for comparison (gray dashed lines). Panels
(e) and (f) show the distributions for the pressure at, respectively, 1.5 and 2.5 $n_{sat}$ at the different stages
(line colours correspond to the ones used in the other four panels) of the analysis, with the combined
multi-messenger+HIC region shaded in light-blue. See \cite{Huth2022} for more details	
}\label{fig:astro}
\end{figure}

Recently, the pioneering work of \cite{Huth2022} has gathered astronomers and nuclear physicists to combine astrophysical multi-messenger observations of neutron stars 
with HIC data from the FOPI and ASY-EOS experiments within a Bayesian framework to better constrain the dense-matter EoS. 
The analysis starts from a prior EoS derived from local chiral EFT interactions (Figure~\ref{fig:astro}a), which exhibits large uncertainties at high densities. 
These uncertainties are progressively reduced by including astrophysical constraints from precise mass measurements of the massive neutron stars PSR J0348+0432 
and PSR J1614-2230, X-ray pulse-profile modelling of PSR J0030+0451 and PSR J0740+6620 obtained by NICER and XMM-Newton,
and gravitational-wave observations of the mergers GW170817 and GW190425 (Figure~\ref{fig:astro}b).
Additional constraints from HIC data are then imposed (Figure~\ref{fig:astro}c), and the final EoS is obtained by combining astrophysical and HIC information (Figure~\ref{fig:astro}d). 
This analysis yields a neutron-star radius of $12.01^{+0.37}_{-0.38}$~km (at $\approx 68\%$ confidence) for a 1.4 solar mass star. 
The inclusion of HIC constraints increases slightly the pressure around $1.5 n_{sat}$ (see Figure~\ref{fig:astro}e), 
where the sensitivity of HIC observables is largest, leading to slightly larger neutron-star radii consistent with recent NICER results. 
At higher densities ($\approx2.5 n_{sat}$, see Figure~\ref{fig:astro}f), 
the pressure is mainly constrained by astrophysical observations, reflecting the reduced sensitivity of current HIC measurements in this region. 
These findings highlight the complementarity between HIC experiments and multi-messenger observations in constraining nuclear matter at intermediate densities. 
They also underline the importance of reducing experimental uncertainties and developing more model-independent analyses of HIC data, 
particularly to better constrain the symmetry energy above $1.5 n_{sat}$, 
a density region crucial for neutron-star modelling.

\begin{figure}[!htb]
\centering
\includegraphics[width=0.98\textwidth]{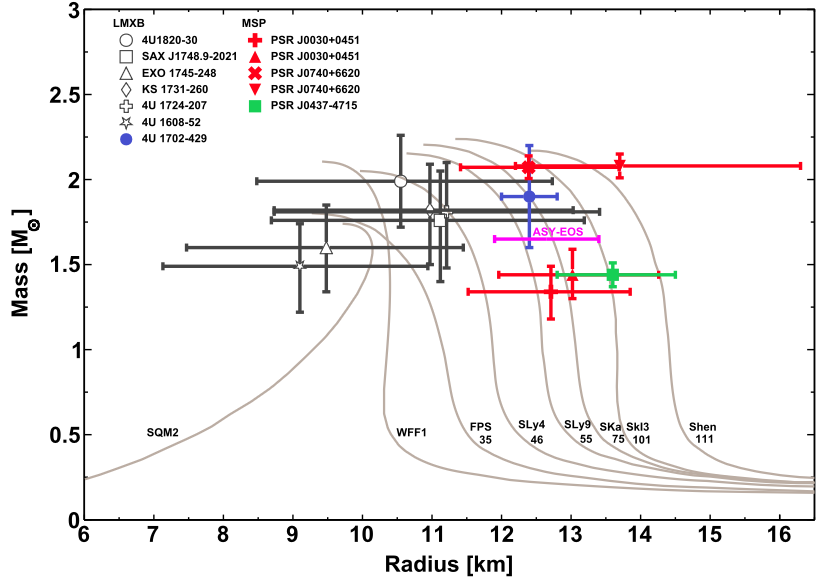}
\caption{(Reprinted from \cite{Russotto2023}) 
Masses and radii for selected individual neutron stars. LMXB data have been obtained from the observation of low-mass X-ray binary systems.
MSP (millisecond pulsars) data, generally of better precision, have been provided by a new generation of instruments such as NICER.
The radius range labeled as "ASY-EOS" was obtained by translating
the symmetry energy slope parameter (L) range obtained from the ASY-EOS heavy-ion experiment to the
neutron star radius range \cite{Trautmann2019}. It corresponds to a neutron star mass of 1.4 solar mass, 
but has been displaced arbitrarily in vertical direction for clarity. 
The labeled gray lines represent the mass-radius relations for selected EoSs. 
The numbers below the acronyms
of the EoSs denote the corresponding values of the symmetry energy L parameter in MeV. 
More details are given in \cite{Russotto2023}.
}\label{fig:NSMR}
\end{figure}

In the review \cite{Russotto2023}, the same HIC constraints were used to derive predictions for the neutron-star mass–radius relation (see Figure.~\ref{fig:NSMR}), 
showing good agreement with the most recent and precise astrophysical measurements, with a comparable -- if not superior -- level of precision.

A related study by \cite{Ghosh2022} also employed ASY-EOS data to investigate multi-physics constraints on the neutron-matter EoS. 
In this work, the prior EoS was obtained within a relativistic mean-field (RMF) approach, 
while chiral EFT calculations at low densities and multi-messenger observations at high densities were used as constraints to determine the posterior distribution. 
HIC results from the KaoS, FOPI and ASY-EOS experiments were included as intermediate-density constraints but treated separately because of their model dependence.
Requiring agreement with ASY-EOS data in the range $1.1-2.0 n_{sat}$ selects about $40\%$ of the prior EoSs. 
The resulting posterior distributions with and without HIC constraints show only moderate differences, 
although the authors emphasise that more precise and model-independent HIC measurements could significantly improve EoS constraints in the $1-2 n_{sat}$ region.

Comparing the two approaches, one finds that Ref.~\cite{Huth2022} starts from a softer chiral EFT-based EoS up to $1.5 n_{sat}$, 
which becomes stiffer once HIC constraints are included. 
And in contrast to \cite{Huth2022} where the proton fraction is fixed by the $\beta$-equilibrium condition 
-- assuming that densities of electrons and protons are equal due to local charge neutrality --, the RMF model adopted by Ghosh et al. imposes a more isospin-symmetric matter at high densities, 
making the resulting EoS less sensitive to the symmetry-energy contribution.

Although not exhaustive, these multi-physics analyses illustrate how HIC measurements -- such as those from the FOPI and ASY-EOS experiments -- 
provide laboratory constraints on asymmetric nuclear matter that complement astrophysical observations. 
Their combined use can significantly improve EoS determinations, 
with future progress to be expected from more model-independent analyses and new high-precision experiments probing densities around $2n_{sat}$.

\subsection{Nuclear theory and observations}\label{sec:comparison2}

\begin{figure}[!htb]
\centering
\includegraphics[width=0.7\textwidth]{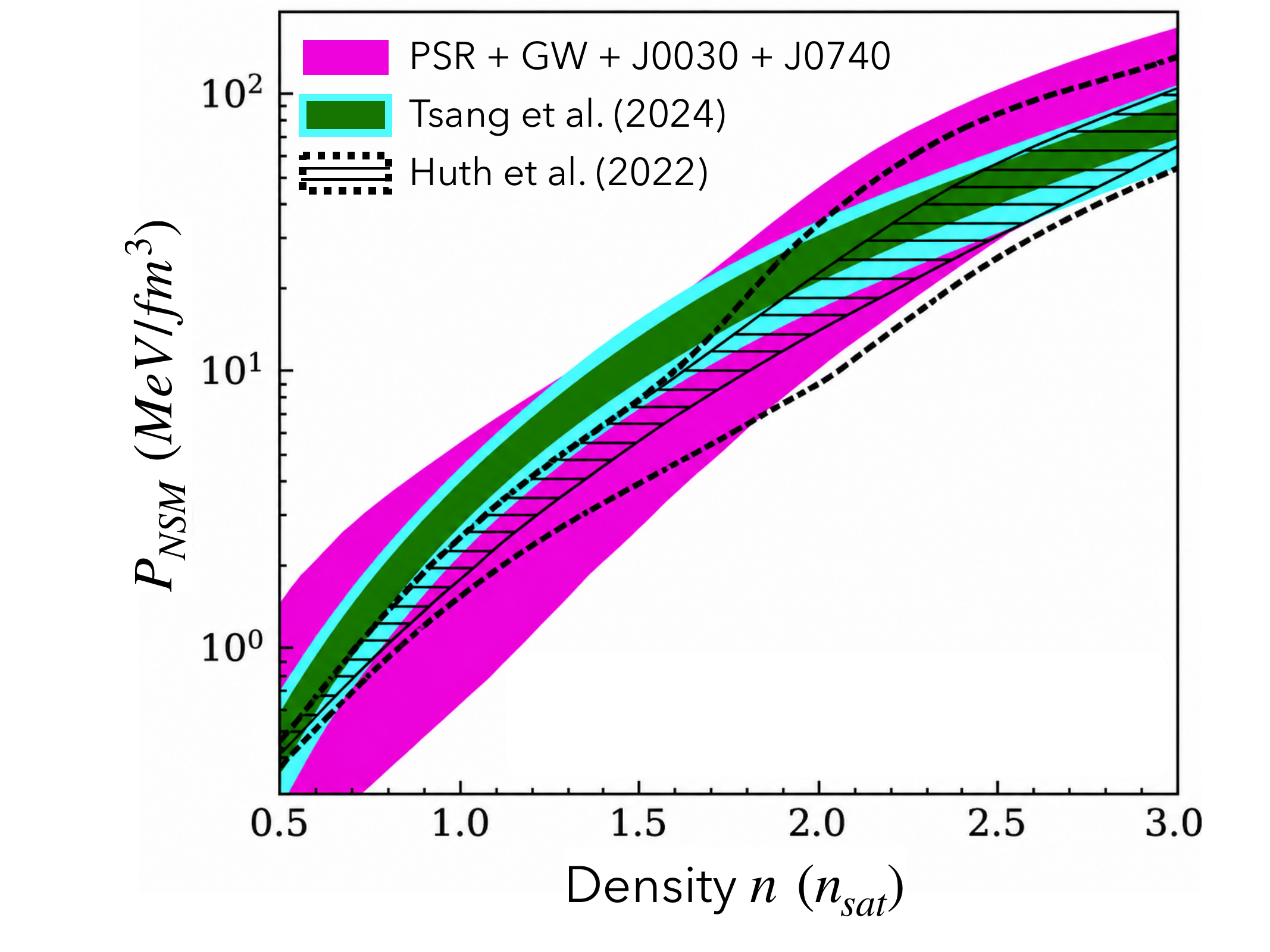}
\caption{(Adapted from \cite{Tsang2024}) 
Constraint of the pressure in neutron-star matter from various Bayesian inferences. The magenta area depicts the EoS obtained in \cite{Legred2021} from
astronomical constraints using only a non-parametric EoS. The hatched contour represents the posterior of  \cite{Huth2022} obtained with astronomical and high-density HIC (FOPI and ASY-EOS) constraints at high energy and chiral EFT as priors at low density (given by the dotted lines as in Figure~\ref{fig:astro}). The dark green and light blue shaded regions represent respectively
$68\%$ and $95\%$ confidence boundaries of the posterior distributions obtained in \cite{Tsang2024} using a combination of 15 constraints composed of three astronomical observations and
12 nuclear experimental constraints.
}\label{fig:tsang}
\end{figure}

Recently, similarly to \cite{Ghosh2022}, authors of \cite{Tsang2024} have warned about the use of the chiral EFT approach 
to build the prior of Bayesian analysis like adopted in \cite{Huth2022}. 
They have emphasised the strong tension between the extrapolation in neutron stars of the chiral EFT pressure between $1-1.5 n_{sat}$, for which the pressure distribution 
is situated at lower values than what HIC and astronomical observation indicate. Therefore, it may create a bias towards an artificial softening of the pressure 
in this density region, as illustrated by Figure~\ref{fig:tsang} derived from this work. 
	What is shown in this figure and in the similar Figure 4.2 in \cite{Sorensen2024} is introduced and discussed in detail in \cite{Cozma:2025dyp} (Section~8 and Fig.~10).

\section{Towards higher density and higher precision constraint of the EoS with HIC}\label{sec:perspectives}

\subsection{Symmetric nuclear matter EoS}\label{sec:perspectives1}

\begin{figure}[!htb]
\centering
\includegraphics[width=0.49\textwidth]{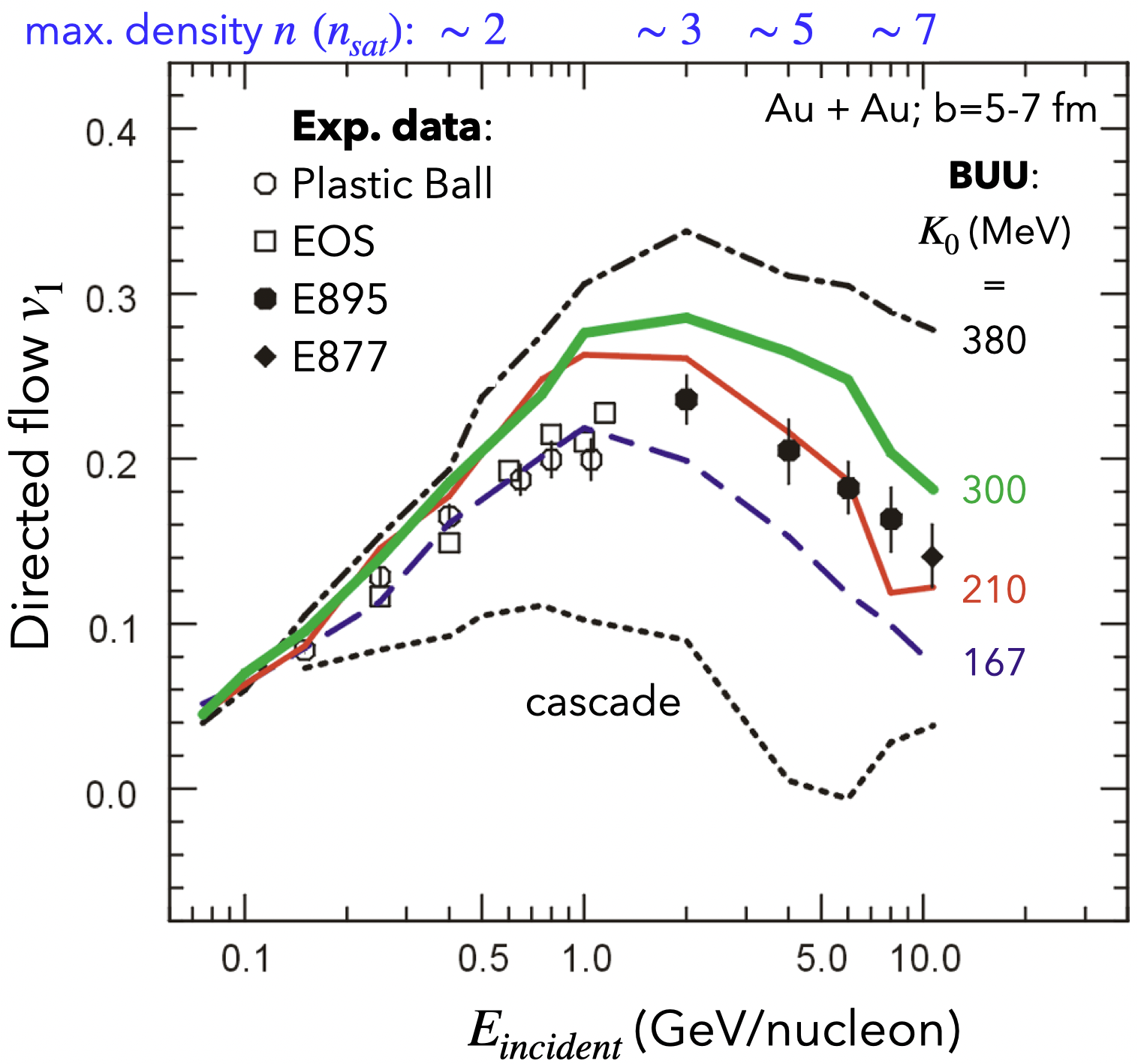}
\includegraphics[width=0.50\textwidth]{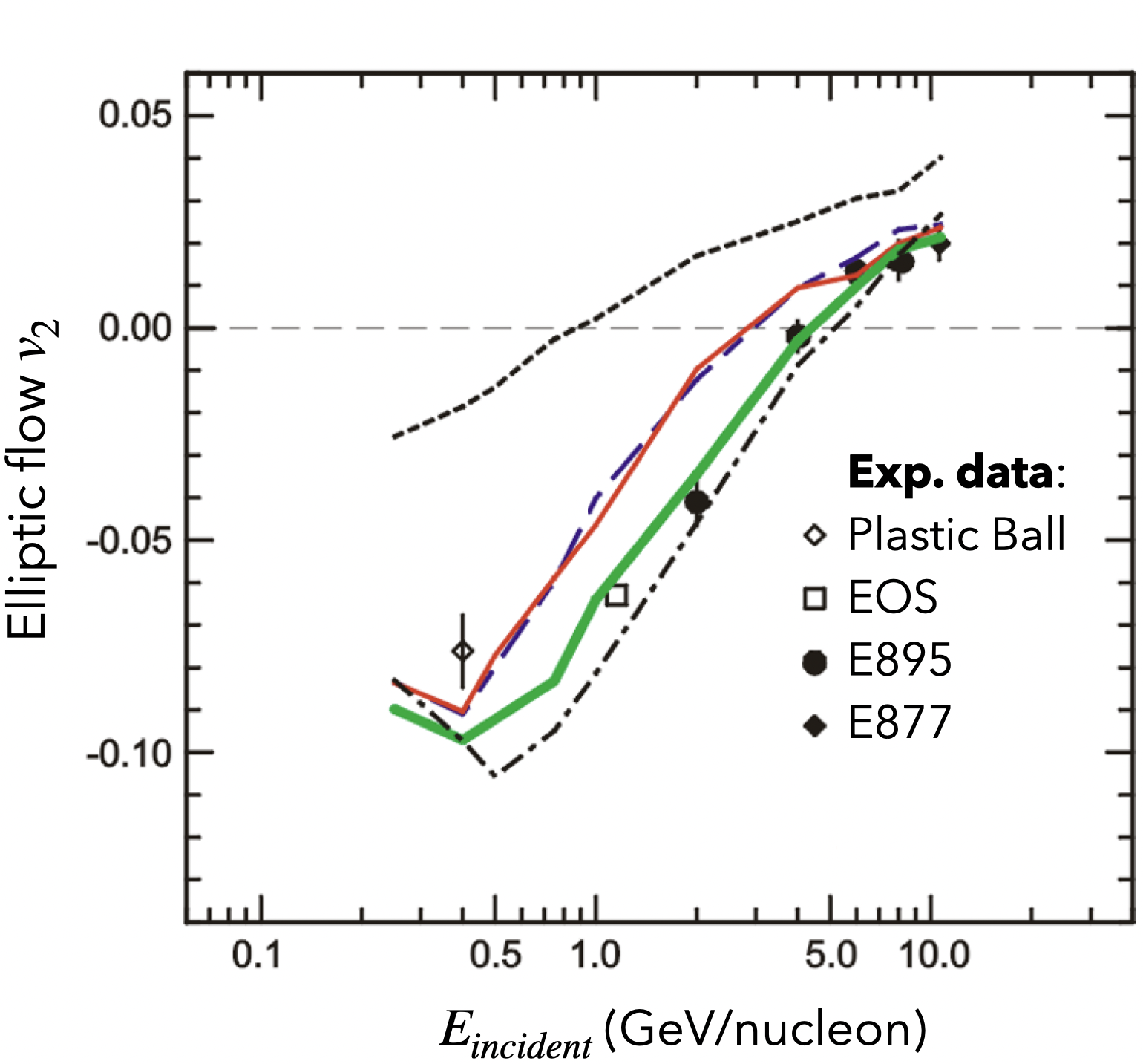}
\caption{(Adapted from \cite{Danielewicz2002}) 
Left panel: SNM EoS constraint from directed flow measured at mid-rapidity in semi-central \AuAu collisions as a function of the incident energy per nucleon, 
from several experiments performed at AGS and Bevalac. 
The solid and open points show experimental values. Lines display cascade and BUU transport theory predictions with various stiffnesses 
indicated by their $K_0$ value in MeV. The top scale $n$ indicates the maximum density in $n_{sat}$ units reached in collisions according to the theory.  
Right panel: same representation with the elliptic flow. 
}\label{fig:science2002}
\end{figure}

As we have seen, the SNM EoS has been accurately constrained so far up to $\approx 3n_{sat}$ at GSI. 
At higher densities, at the moment, very little is available in the literature as constraints, apart from the work of \cite{Danielewicz2002}, 
using AGS and Bevalac flow data of \AuAu collisions to 
deduce an EoS constraint, but with a lower accuracy than that achieved at GSI. In this work, reached densities are $\approx 3-5 n_{sat}$. 
This work confirms expectations of \cite{ALF2018} that above few GeV/nucleon incident energy, the elliptic flow begins to be very little constraining, 
whereas the proton directed flow (noted $v_1$ or F) is to be considered instead as shows Figure~\ref{fig:science2002} extracted from this work.
However, $v_2$ data points used in this work show differences with those measured by FOPI and HADES at GSI at the same energies  which questions 
the accuracy of the analysis, and could explain the inconsistency reported in this study between the soft EoS favoured by $v_1$ at the higher beam energies, 
and the stiff EoS favoured by $v_2$ at the lower energies. 
Consequently, priority should be given to repeating this analysis, firstly using corrected and supplemented $v_1$ and $v_2$ flow values, 
incorporating the new \AuAu data from FOPI, HADES and BES, 
and secondly using the same pBUU transport code in its updated version, as well as other QMD-type transport models, in order to better identify systematic errors.  

Furthermore, future experiments like CBM and HADES at FAIR (Facility for Antiproton and Ion Research, Germany), and BM@N at NICA (Nuclotron-based Ion Collider fAcility at the Joint Institute for Nuclear Research, Russia) should be able to contribute to this systematics with even more precise flow data.  	

\begin{figure}[!htb]
\centering
\includegraphics[width=0.7\textwidth]{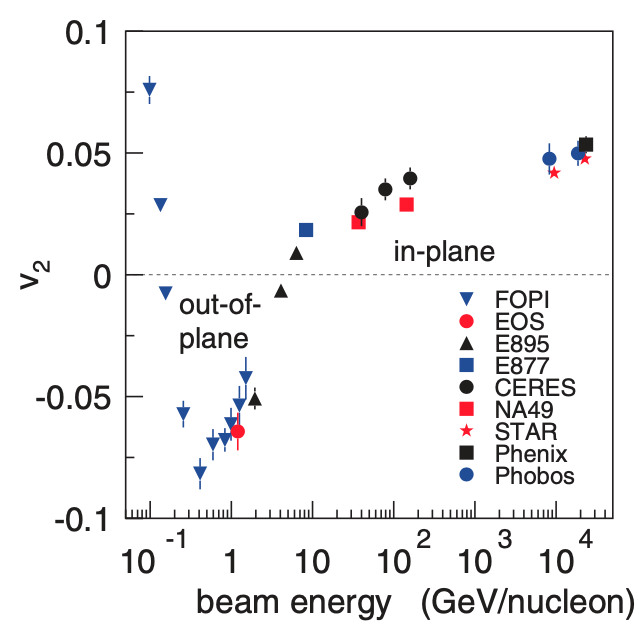}
\caption{(Reprinted from \cite{ALF2018}) 
Elliptic flow $v_2$ of $Z=1$ particles at mid-rapidity as a function of incident beam energy in semi-central \AuAu collisions as measured by various experiments (see references in \cite{ALF2018}), indicated by 
different symbols.
}\label{fig:ALF2018}
\end{figure}

As demonstrated by \cite{Danielewicz2002, ALF2018}, and illustrated by Figure~\ref{fig:ALF2018}, 
the elliptic flow as a method to constrain the EoS finds its limits above few GeV per nucleon bombarding energy, because then, particles start to be preferentially emitted in-plane,
as they are no longer shadowed by spectators which are escaping too fast with regard to the expansion speed of the fireball. 
At intermediate beam energies (0.1-3 GeV/nucleon), out-of-plane emissions are favoured by the screening effect of spectators on the expansion of the fireball, 
resulting in a negative $v_2$. The stiffer the EoS, the faster the fireball expands, the stronger is the spectator shadowing effect, consequently the more negative is $v_2$. 
Above 3 GeV/nucleon, $v_1$ must be considered instead as an EoS sensitive probe.

Other harmonics of the flow (see Equation~\ref{eq:flow}) may provide useful constraints on the stiffness of the EoS.
In particular, in the recent review \cite{Sorensen2024}, the triangular flow $v_3$ is predicted to be a good candidate as supported by predictions of the SMASH transport model.

\subsection{Symmetry energy}\label{sec:perspectives2}

As seen in Section~\ref{sec:Esym}, at the moment, the rather precise knowledge of the symmetry energy does not go beyond $\approx 1.5 n_{sat}$, 
which represents the main limitation so far of HIC in addressing the EoS of neutron-star matter at high densities. 
As seen in Section~\ref{sec:comparison1}, in particular on Figure~\ref{fig:astro}f, above $2n_{sat}$, the posterior distribution of the pressure in a neutron star is primarily driven by astronomical observations because reliable HIC data concerning the symmetry energy are not yet available at higher densities.

A second campaign of measurements of the ASY-EOS Collaboration, still based on flow measurement in \AuAu collisions, has been carried out at GSI in Spring 2025, with the purpose to push this frontier towards higher densities 
($\approx 2.5 n_{sat}$) and improve the accuracy of the constraint on $E_{sym}$.  To do so, as regard to ASY-EOS I campaign \cite{Russotto2016}, more and higher beam energies
 (280, 400, 600, 1000 MeV/nucleon) have been measured, 
with improved statistics, and a new set-up allowing a better resolution of reaction plane, neutrons and protons, and slightly larger acceptance. 
The data analysis is still on-going. More details on the ASY-EOS II experiment are given in \cite{Russotto2023}. 

Many facilities in the world offer perspectives of precising the knowledge of the symmetry energy above saturation density. On its side, the higher beam energies and increased rare-isotope capabilities provided by FRIB400 (Facility for Rare Isotope Beams, Michigan State University) would enable as well heavy-ion collisions to probe nuclear matter at densities above saturation density. By measuring observables such as neutron/proton collective flows, pion and light-isobar production, and isospin transport with improved precision in highly neutron-rich systems, FRIB400 is expected to significantly reduce uncertainties on the high-density behavior and momentum dependence of the symmetry energy \cite{Brown2024,FRIB400_2019}. Other already existing facilities may help complementing the knowledge of the symmetry energy at near/supra-saturation density like RIKEN (Japan) and GANIL (France). RIKEN offers strong prospects for studying the symmetry energy at supra-saturation density through high-intensity radioactive ion beams and upgraded detector systems enabling precision measurements of neutron/proton flows, pion production, and other isospin-sensitive observables in neutron-rich heavy-ion collisions up to a few hundred MeV per nucleon, providing sensitive probes of the symmetry energy through observables such as isospin diffusion, nucleon collective flows, and light-particle production. The intermediate-energy heavy-ion beams available at GANIL can compress nuclear matter to around saturation density and moderately above ($\approx 1-2\rho_0$), providing as well sensitive probes of the symmetry energy through observables such as isospin diffusion, nucleon collective flows, and light-particle production.

As we have seen in Section~\ref{sec:perspectives1}, other observables than $v_2$ will have to be used 
in order to probe the symmetry energy at higher densities ($>3n_{sat}$) obtained at higher beam energies (few GeV/nucleon). 

\begin{figure}[!htb]
\centering
\includegraphics[width=0.98\textwidth]{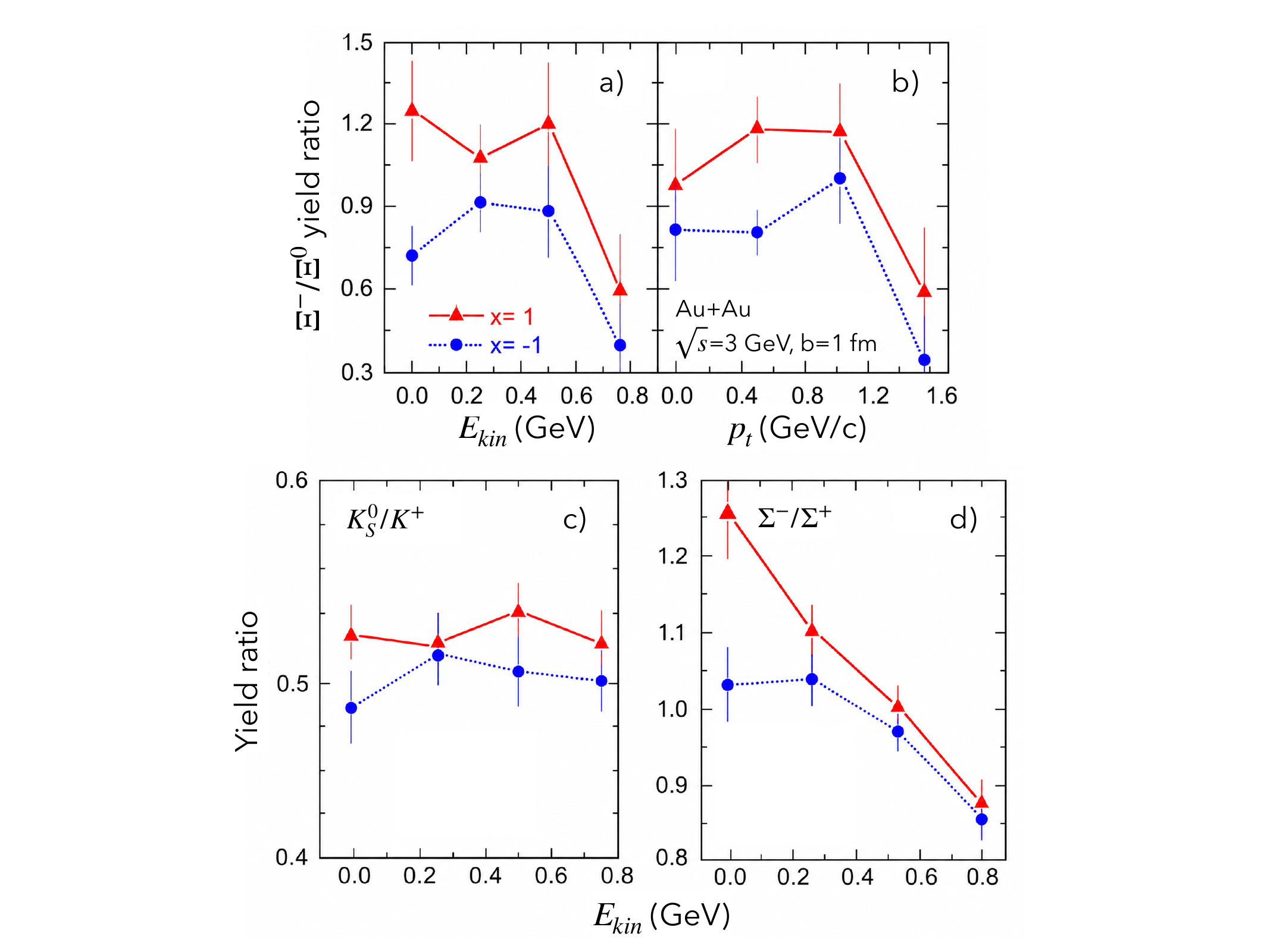}
\caption{(Adapted from \cite{Yong2022}) 
Predictions of particle production in central \AuAu collisions at $\sqrt{s_{NN}}$ = 3 GeV  with the updated ART (A Relativistic Transport)
model with momentum dependent isoscalar and isovector single-nucleon mean-field potentials 
corresponding to different symmetry energies at supra-saturation densities, with stiﬀ (blue circles and dashed lines)
and soft (red triangles and full lines) symmetry energies. 
Panels a and b display respectively kinetic energy and transverse momentum distributions of the doubly strange baryon $\Xi^-/\Xi^0$ yield ratio. 
Panels c and d show kinetic energy distributions of  $K^0_S/K^+$ and $\Sigma^-/\Sigma^+$ respectively.
}\label{fig:yong}
\end{figure}

Several observables have been proposed to probe the symmetry energy at high densities, including isospin-sensitive particle production such as the 
$\pi^-/\pi^+$ ratio \cite{Cozma2024,Greco:2009ck,Xiao2009}, the kaon multiplicity ratio $K^+/K^0_S$ \cite{Ferini2006}, and the yields and flows of neutrons, protons, and light clusters \cite{Russotto2011}. Other isospin multiplets like $\Sigma^-/\Sigma^+$ and $\Xi^-/\Xi^0$  have been more recently proposed 
as potentially sensitive to $E_{sym}$ at FAIR energies by \cite{Yong2022}, additionally to $K^+/K^0_S$ yield ratios, as illustrated by Figure~\ref{fig:yong}.  
Experimentally, $\Sigma$ baryons are difficult to identify because they are short-lived, and decay with at least one neutral daughter particle 
which requests high-resolution tracking and vertexing detectors close to the target to reconstruct the missing mass, 
like planned for the CBM set-up at FAIR which is well situated to exploit such probes in the next decade.

However, compared with constraints on the isoscalar (SNM) part of the EoS, the situation remains less clear. 
For instance, the double-kaon multiplicity ratio, although promising in infinite nuclear matter calculations, showed limited sensitivity to the symmetry energy 
due to limited statistics and to high systematic uncertainties \cite{Lopez2007}. Similarly to \cite{Cozma2024} for pions, using kaon yield ratios probably implies 
that in transport models used to interpret the data, 
$\Delta$ potentials, cross-sections, and momentum dependencies of potentials will have to be well constrained to reduce uncertainties.  
 
Observables based on neutron–proton ratios, such as the differential directed-flow ratio proposed by Bao-An Li \cite{BALi2002}, 
are experimentally challenging because accurate neutron detection over a large solid angle is required. 

Overall, experimental information on $E_{sym}$ at high densities from HIC is still limited compared with the low-density regime. 
This is largely due to experimental difficulties, particularly the need for large-acceptance neutron detectors -- of which only a few exist worldwide -- 
and highly precise pion-tracking systems. 
Consequently, new and more accurate measurements are required to confirm current findings and extend constraints to higher densities.

\subsection{Transport models}\label{sec:perspectives3}

The interpretation of HIC data is generally carried out using semi-classical transport models, 
which allow the study of several properties of nucleon–nucleon interactions in dense matter, including the nuclear EoS, 
the effective nucleon masses, and in-medium scattering cross sections. 
Several classes of models are currently used -- such as BUU, TDHF, BL and QMD. BUU and QMD type of models are presently the two most frequently 
applied to investigations of the EoS.
Obtaining reliable constraints on the EoS requires extensive benchmarking of these models against the large body of available experimental data, 
as well as a detailed understanding of the differences among the various transport approaches. 
Addressing these issues is a central focus of ongoing theoretical work, including efforts within the TMEP Collaboration (Transport Model Evaluation Project) \cite{Xu2016,Zhang2018,Ono2019,Colonna2021,Wolter2022}.
Recent studies have identified several sources of discrepancies in model predictions of pion production in heavy-ion collisions. 
Clarifying these differences is also essential for exploiting cleaner probes of the EoS, such as kaon emission. 
Current efforts are therefore directed toward understanding the role of factors such as the momentum dependence of effective interactions 
and toward establishing benchmark calculations that will enable more reliable 
and precise constraints on the nuclear EoS from both existing and future heavy-ion collision experiments. 
These efforts are key for reducing theoretical uncertainties, hence for improving the accuracy of EoS constraints deduced from HIC data in the light of these models. 

\begin{figure}[!htb]
\centering
\includegraphics[width=0.7\textwidth]{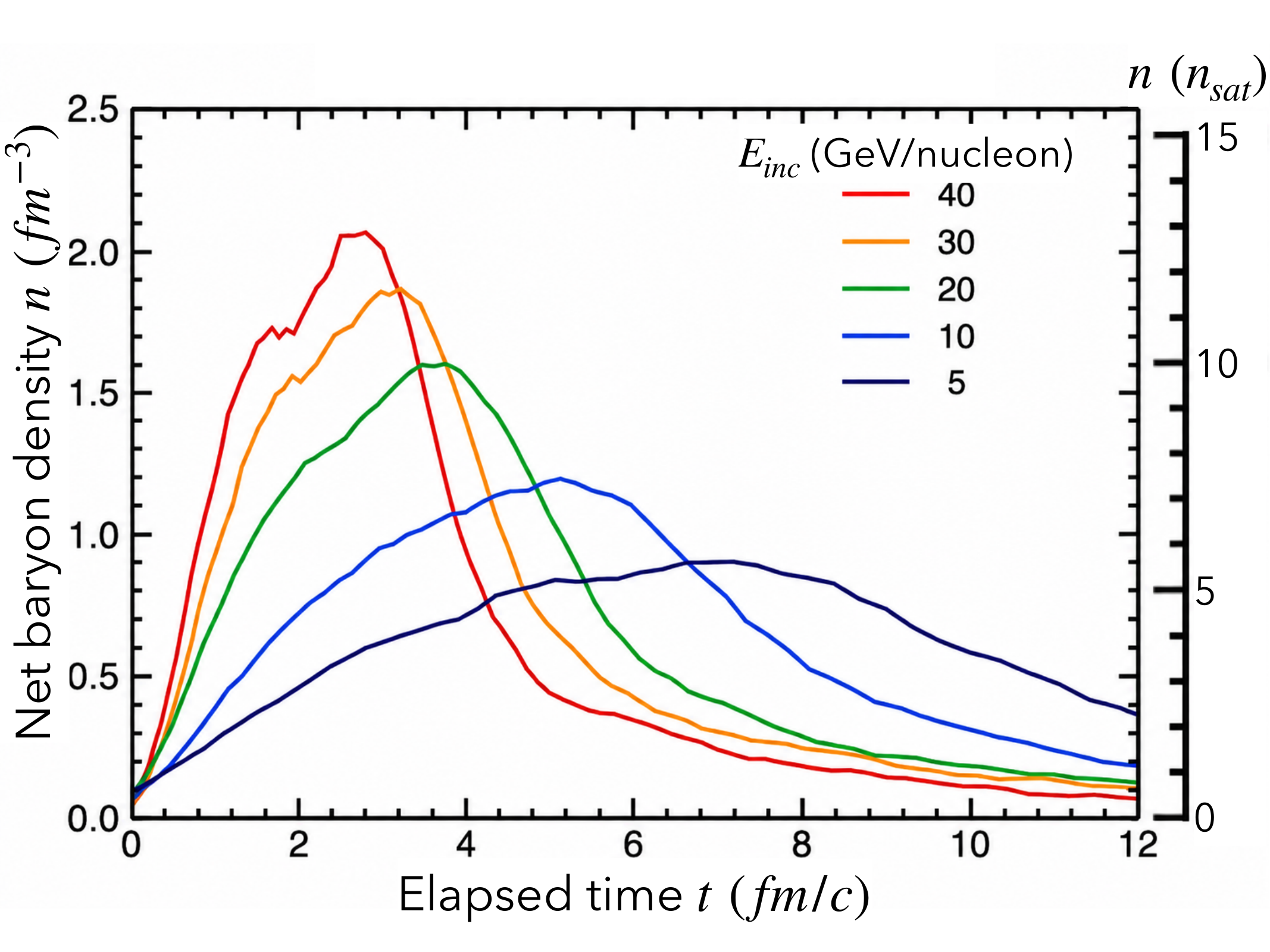}
\caption{(Adapted from \cite{Arsene2007}) 
Time evolution of the central net baryon densiy at the centre of central
\AuAu collisions at various bombarding energies $E_{inc}$, as predicted by the PHSD model.
}\label{fig:phsd}
\end{figure}

\begin{figure}[!htb]
\centering
\includegraphics[width=0.7\textwidth]{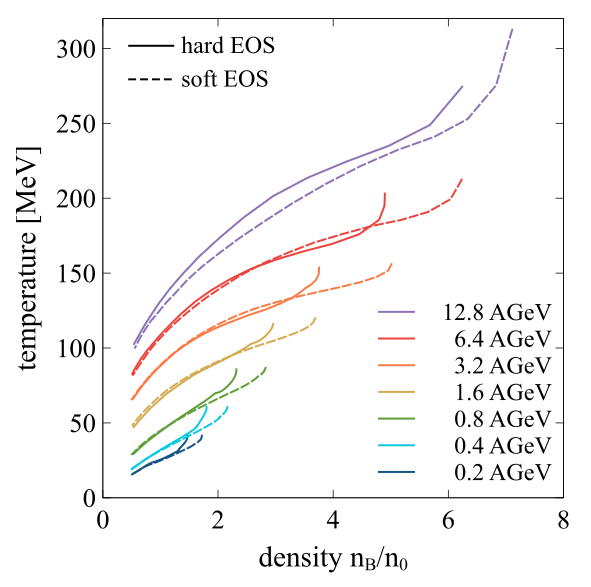}
\caption{(Reprinted from \cite{Sorensen2024}) 
Phase diagram trajectories (in the representation of the temperature as a function of baryon density in units of $n_{sat}$) in the mid-rapidity region of central \AuAu collisions at various incident energies (indicated in the legend with various line colours), as predicted by UrQMD simulations 
with a soft (dashed lines) or a hard (full lines) EoS, corresponding to $K_0=200$ and $380$~MeV respectively. Trajectories are given only 
when the temperature is fairly well-defined, from the highest compression phase up to when densities reach around $0.5 n_{sat}$.
}\label{fig:sorensen}
\end{figure}

The second challenge deals with the perspective of constraining the EOS at larger densities, 
i.e. with higher beam energies (as illustrated by Figure~\ref{fig:phsd}) which are beyond the limit of applicability of many of those models, 
as they do not contain the proper relativistic features, and non-perturbative treatment of particle generation and dynamics. 
Models already adapted to ultra-relativistic energies, on their side, often 
lack of the state-of-the-art features like momentum dependences of potentials, effective-mass corrections that are needed to correctly assess EoS effects. 
Another issue is that high energy HIC unfortunately explore temperatures that are far above conditions found in neutrons star (and mergers), as presented in Figure~\ref{fig:sorensen}.
When temperatures in HIC become very high ($\gtrsim 100$~MeV), extracting the zero-temperature nuclear EoS using mean-field transport models 
becomes difficult for several reasons. 
First, the thermal contributions to the pressure: at high temperatures, a large fraction of the pressure comes from thermal excitations, not only from the cold EoS. 
Observables then reflect a combination of thermal and interaction effects, making it hard to isolate the $T=0$ EoS. 
Second, the breakdown of the mean-field picture: mean-field transport models assume that nucleons move in an average potential. 
At high temperatures and densities, however, frequent collisions, strong fluctuations, and many-body correlations \cite{Colonna2020} become important, reducing the validity of a simple mean-field description, in particular when the low density limit -- according to the Brückner theory -- is exceeded \cite{Hartnack1998}. 
Third, the abundant particle production: temperatures above $\sim100$~MeV lead to strong production of resonances and mesons (e.g., $\Delta$, pions, kaons). 
The system is no longer dominated by nucleons, and the relation between observables and the nucleonic EoS becomes indirect and model dependent. 
Fourth, the uncertain in-medium properties: at such temperatures and densities, quantities like in-medium cross sections, resonance potentials, 
and effective masses are poorly known, introducing additional uncertainties in transport calculations.
For these reasons, HIC at moderate incident energies, where temperatures are lower and nucleonic dynamics dominate, generally provide cleaner constraints on the cold nuclear EoS.
 
Overall, the main goal of the field in a long-term perspective, should be providing a common EoS with well-defined errors consistent with all transport codes, as advocated in 
\cite{Sorensen2024}, a recent collaborative review on the topic of transport modelists and experimentalists  
aiming at summarising and coordinating current efforts to determine the nuclear equation of state (EoS) using heavy-ion collision experiments.
Such initiatives, along with TMEP, should be encouraged in order to: 
1- review how heavy-ion collisions probe dense nuclear matter, reaching baryon densities up to several times nuclear saturation density and temperatures up to well above 100~MeV.
2- discuss the experimental observables (e.g., particle production and collective flow) that can constrain the EoS and its dependence on density, temperature, and isospin asymmetry.
3- evaluate the role and limitations of hadronic transport models used to interpret heavy-ion data and extract EoS information.
4- highlight the connections between heavy-ion experiments, microscopic nuclear theory, and neutron-star observations, emphasising a multi-messenger approach to dense-matter physics.

\subsection{Strangeness in neutron stars}\label{sec:perspectives4}

\begin{figure}[!htb]
\centering
\includegraphics[width=0.75\textwidth]{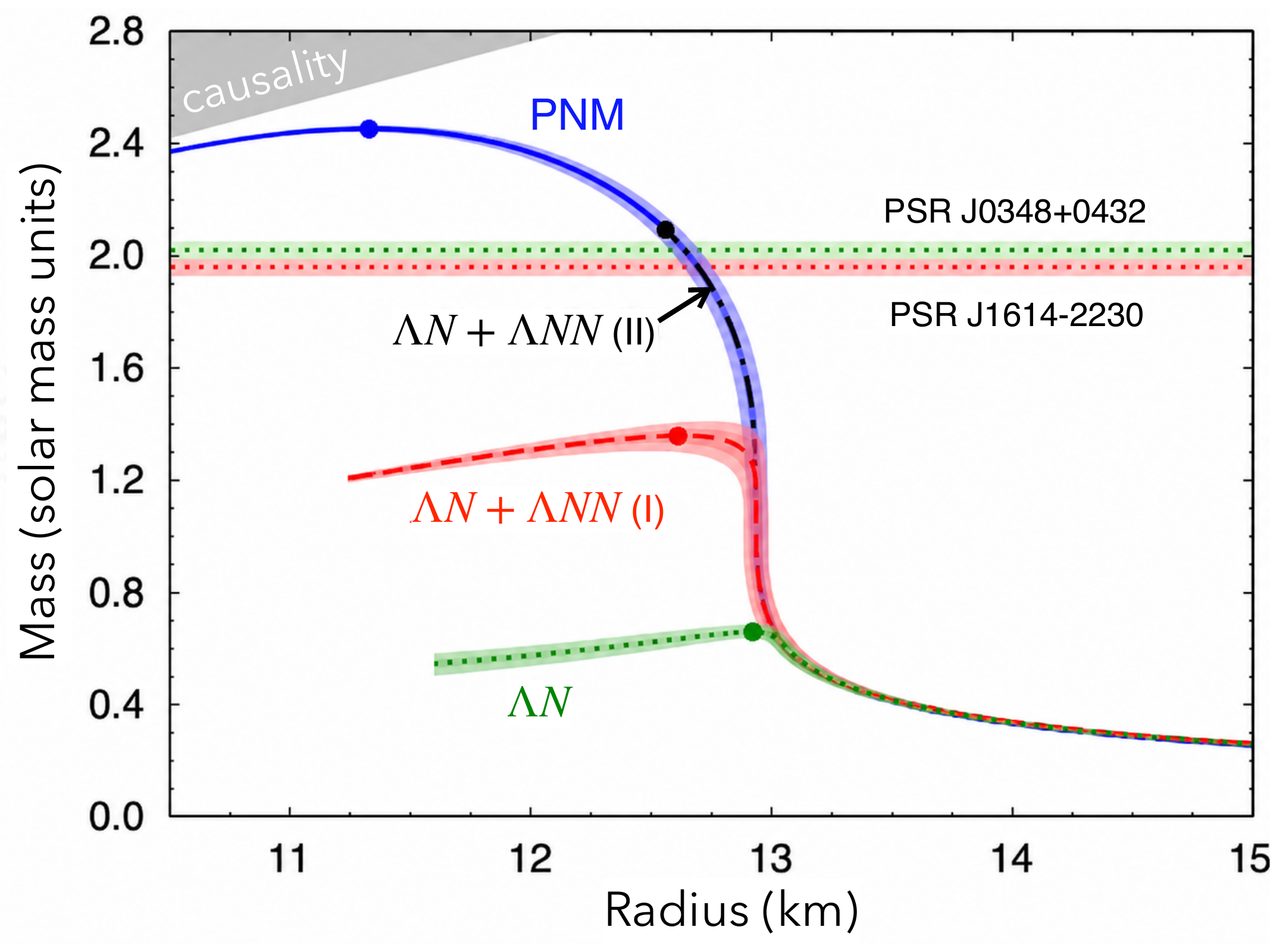}
\caption{(Adapted from \cite{Lonardoni:2014bwa}) 
Neutron-star mass-radius relations according to various ansatz of the hyperon-nucleon interaction.
The blue solid curve refers to a pure neutron matter EoS. 
The green dotted curve represents the EoS of hyperonic matter
with hyperons interacting via only a two-body $\Lambda N$ force.
The red and black dashed curves are obtained including the three-body
$\Lambda NN$ potential from two different models (I and II). Shaded areas represent the uncertainties.
Full circles on these curves indicate the corresponding predicted maximum masses. 
The two horizontal bands at about 2 solar mass show observed masses of the heavy pulsars PSR J1614-2230
and PSRJ0348+0432. The region excluded by causality is depicted by the grey shaded area.
}\label{fig:lonardoni}
\end{figure}

Constraining the density dependence of the hyperon–nucleon ($Y-N$) interaction is a key objective for both nuclear physics and astrophysics, 
as it directly impacts our understanding of neutron star interiors. At sufficiently high baryon densities, the chemical potentials of nucleons 
become large enough to make the appearance of strange baryons (such as $\Lambda$, $\Sigma$, $\Xi$ hyperons) energetically favorable, 
as long as only Y-N 2-body forces are considered \cite{Djapo:2008au,Lonardoni:2014bwa}. 
The onset and population of these hyperons are governed by the poorly known $Y-N$ and $Y-Y$ interactions, 
especially their behavior with increasing density. Moreover, according to D. Lonardoni \cite{Lonardoni:2014bwa}, 
the knowledge of three-body forces is essential as well.
Since hyperons introduce additional degrees of freedom, their emergence generally softens the EoS, reducing the pressure at a given density. 
This has major consequences for neutron stars, in particular for their maximum mass and radius. 
The so-called “hyperon puzzle” arises from the apparent contradiction between this softening and the observation of neutron stars with masses around two solar masses, 
which require a sufficiently stiff EoS. However, Bednarek et al. \cite{Bednarek2012}, and later confirmed by Lonardoni et al. \cite{Lonardoni:2014bwa}, 
have demonstrated that the appearance of hyperons does not necessarily cause a puzzle,
when three-body hyperon-nucleon forces, playing an important role, are taken into account, as illustrated by Figure~\ref{fig:lonardoni}. 
According to \cite{Lonardoni:2014bwa}, the consequence is that an EoS constrained on binding energies of light hypernuclei is not suﬃcient 
to draw any definite conclusion about the composition of the core of neutron stars.

Experimental constraints on the $Y–N$ potential, and especially on its density dependence, are therefore essential to resolve this puzzle. 
Information at or near nuclear saturation density can be obtained from hypernuclei, which provide relatively well-established constraints on the depth of the $\Lambda-N$ potential. 
However, neutron star cores probe much higher densities, making it crucial to determine how these interactions evolve beyond saturation density.
The most relevant density range for neutron star studies lies about $2-4 n_{sat}$, where hyperons are expected to first appear and significantly influence the EoS \cite{Bednarek2012,Jangal2025}. 
In this regime, even modest uncertainties in the $Y–N$ interaction can lead to large differences in predicted neutron star properties.
HIC and future dedicated experiments offer a promising avenue to access this information. 
Observables related to hyperon production, flow, and correlations in dense matter can provide indirect constraints on in-medium $Y–N$ potentials. 
Combined with theoretical developments in many-body approaches and constraints from neutron star observations, 
these experimental efforts are crucial for establishing a consistent picture of dense matter. 

The extraction of information on the density dependence of the hyperon–nucleon interaction from HIC relies critically on transport models 
with accurate parametrizations of in-medium interactions and particle production channels, in particular those involving strangeness 
(e.g. kaon production), which are known to be sensitive probes of the high-density EoS (see \cite{Hartnack2012}).

Ultimately, improving our knowledge of the density dependence of hyperon interactions is indispensable for determining 
whether hyperons are present in neutron star cores and for constructing a realistic high-density EoS. 
According to a recent study of Bauswein et al. \cite{Bauswein:2025dfg}, it is expected that once the next generation of gravitational-wave telescopes (like the Einstein Telescope) is online, 
with required precision for measurements of the tidal deformability of high-mass neutron stars of a few per cent, 
experimental signatures of strangeness in neutron stars will have a fair probability to be found. 

\subsection{QCD phase transition in neutron stars}\label{sec:perspectives5}

The possible existence of a phase transition from hadronic matter to deconfined quark matter is one of the central open questions in dense nuclear physics, 
with direct implications for neutron star structure and dynamics. 
In the cores of massive neutron stars, baryon densities can reach several times nuclear saturation density, where nucleons may no longer remain the relevant degrees of freedom. 
Instead, matter could transition to a phase of deconfined quarks, possibly including exotic states such as color-superconducting phases \cite{Alford1999,Sedrakian2019,Gholami2025}. 
Identifying signatures of such a QCD phase transition is therefore crucial for establishing the true composition of neutron star interiors.
The presence of a phase transition would significantly affect the equation of state (EoS), 
typically leading to a softening of the pressure–density relation at the transition, 
followed at higher densities by a possible stiffening depending on the properties of quark matter. 
This has observable consequences for neutron stars: it can modify the mass–radius relation, limit the maximum mass, 
and even give rise to characteristic features such as “twin stars” (two stars with the same mass but different radii). 
In dynamical scenarios, such as neutron-star mergers, a phase transition could influence the post-merger gravitational-wave signal, 
the collapse time -- in the event that the critical mass is exceeded --, and the emitted neutrino fluxes.
Current theoretical estimates place the onset of a QCD phase transition in neutron star matter at baryon densities of roughly $2–5 n_{sat}$,
 although this range remains highly uncertain and model dependent \cite{Kojo2016, Annala2020, Jokela2019}.
Some models predict an earlier onset, while others push it to higher densities beyond those reached in stable neutron stars. 
Consequently, constraining this transition requires combining inputs from nuclear theory, heavy-ion collisions (which probe high densities at finite temperature), 
and astrophysical observations such as precise mass and radius measurements, as well as gravitational-wave data.
Even though no clear multimessenger observational evidence has so far supported the presence of such a phenomenon in neutron stars \cite{Huth2022,Brandes:2023hma}, 
some studies suggest that the cores of the most massive stars may already be compatible with quark matter \cite{Annala:2023cwx}. 
In this context, searching for signals of a QCD phase transition remains essential for connecting microscopic QCD predictions with macroscopic neutron star observables, 
and for achieving a unified description of strongly interacting matter across a wide range of densities and temperatures.

\section{Conclusions}\label{sec:conclusions}

Heavy-ion collisions at intermediate energies provide a unique laboratory for studying the nuclear EoS under extreme conditions of density, 
temperature, and isospin asymmetry. 
Over the past decades, experimental measurements combined with transport-model simulations have enabled significant progress in constraining 
both the isoscalar part of the EoS and the density dependence of the symmetry energy. 
In particular, observables related to collective flow and particle production -- such as elliptic and directed flow, kaon production, and pion ratios -- 
have proven to be sensitive probes of the pressure generated in compressed nuclear matter.

Results from heavy-ion experiments at intermediate energies indicate that nuclear matter can reach densities of about two to three times the saturation density during the most compressed stage of the collision. 
Constraints derived from flow measurements and sub-threshold particle production suggest a moderately soft EoS at supra-saturation densities in symmetric nuclear matter. 
When combined with constraints on the symmetry energy obtained from observables such as neutron-to-proton (or charged-particle) elliptic-flow ratios, these studies allow estimates of the pressure of neutron-rich matter at densities relevant for neutron stars. 
Remarkably, the pressure–density relations inferred from heavy-ion collisions are broadly consistent with astrophysical constraints derived from neutron-star observations, including gravitational-wave measurements and X-ray studies.

Despite these advances, important challenges remain. The extraction of EoS constraints relies strongly on semi-classical transport models, 
whose predictions can differ because of variations in the treatment of mean fields, in-medium cross sections, and particle production mechanisms. 
Achieving reliable constraints therefore requires systematic benchmarking and a detailed understanding of model dependencies. 
Collaborative efforts such as the Transport Model Evaluation Project aim to reduce these uncertainties and establish common standards for transport calculations.

Another limitation arises from the experimental side. 
Measurements sensitive to the symmetry energy at high densities often require accurate detection of neutrons or precise pion tracking, which are experimentally demanding. 
Consequently, present constraints on the symmetry energy above saturation density remain relatively scarce and less precise than those at lower densities.

Future experimental programs and improved detectors are expected to significantly enhance the situation. 
New campaigns at facilities such as FAIR and NICA with few GeV/nucleon beam energies and upgraded measurements of flow observables and particle production will provide more precise data 
and extend the density range probed in heavy-ion collisions. On its side, the FRIB400 project should be able to deliver precise constraints on the symmetry energy at intermediate incident energies. 
In parallel, developments in transport theory and nuclear many-body calculations will be essential to reduce systematic uncertainties 
and improve the interpretation of experimental observables.
Overall, heavy-ion collision studies play an important role in the emerging multi-messenger approach to dense-matter physics. 
By bridging laboratory experiments with astrophysical observations of neutron stars and their mergers, 
they contribute to building a consistent description of the nuclear EoS across a wide range of densities and physical conditions.



\bibliography{sn-article_v3}



\end{document}